\documentstyle[12pt,graphicx]{article}
\def\beqar {\begin{eqnarray}}
\def\eeqar {\end{eqnarray}}
\def\beq {\begin{equation}}
\def\eeq {\end{equation}}
\def\A{{\cal A}}
\def\B{{\cal B}}
\def\C{{\cal C}}
\def\F{{\cal F}}

\def\S{{\cal S}}

\def\P{{\cal P}}

\def\N{{\cal N}}

\def\al{\alpha}
\def\bt{\beta}
\def\del{\delta}

\def\ga{\gamma}

\def\ep{\epsilon}
\def\la{\lambda}

\def\om{\omega}
\def\Om{\Omega}
\def\th{\theta}
\def\et{\eta}
\def\si{\sigma}

\def\d{\partial}

\def\Ad{{\dot A}}
\def\Bd{{\dot B}}

\def\bz{{\bar z}}
\def\bu{{\bar u}}

\def\bth{{\bar \theta}}

\def\hf{\frac{1}{2}}

\def\<{\langle}
\def\>{\rangle}

\def\Tr{{\rm Tr}}
\def\Path{{\rm P}}

\def\cp{{\bf CP}}
\begin{document}

\begin{titlepage}
\null\vspace{-62pt} \pagestyle{empty}
\begin{center}
\vspace{1.0truein}

{\Large\bf Holonomies of gauge fields in twistor space 1: \\
\vspace{.36cm}
\hspace{-.4cm}
bialgebra, supersymmetry, and gluon amplitudes} \\

\vspace{1.0in} {\sc Yasuhiro Abe} \\
\vskip .12in {\it Cereja Technology Co., Ltd.\\
3-1 Tsutaya-Bldg. 5F, Shimomiyabi-cho  \\
Shinjuku-ku, Tokyo 162-0822, Japan}\\
\vskip .07in {\tt abe@cereja.co.jp}\\
\vspace{1.3in}
\centerline{\large\bf Abstract}
\end{center}

\noindent
We introduce a notion of holonomy in twistor space
and construct a holonomy operator by use of
a spinor-momenta formalism in twistor space.
The holonomy operator gives a monodromy representation
of the Knizhnik-Zamolodchikov (KZ) equation, which is
mathematically equivalent to a linear representation of a braid group.
We show that an S-matrix functional for
gluon amplitudes can be expressed in terms of a
supersymmetric version of the holonomy operator.
A variety of mathematical and physical concepts,
such as integrability, general covariance, Lorentz invariance
and Yangian symmetry, are knit together
by the holonomy operator.
These results shed a new light on
gauge theories in four-dimensional spacetime.

\end{titlepage}
\pagestyle{plain} \setcounter{page}{2} 

\section{Introduction}

In 1988, Nair made a remarkable observation that certain gluon amplitudes
can be interpreted as a current correlator of a Wess-Zumino-Witten (WZW) model
which is to be defined in an extended twistor space \cite{Nair:1988bq}.
There are two important concepts in this observation.
Firstly, by use of spinor momenta, one can
carry out purely helicity-based calculations, {\it i.e.}, calculations
in which gauge fields are dependent solely on the helicity and the numbering
index of each particle, in order to obtain
the so-called maximally helicity violating (MHV) amplitudes \cite{Parke:1986gb}.
It is then naturally required to incorporate $\N = 4$ supersymmetry
in the underlining twistor space to realize a particular form of the MHV amplitudes.
One of the advantages to use a spinor-momenta formalism in twistor space
is that we can obtain a manifestly Lorentz-invariant theory.
This feature is desirable if one tries to construct a theory with general covariance.

The second concept of importance, partly related to the first one, is
that the structure of four-dimensional gluon amplitudes is essentially encoded by a
two-dimensional conformal field theory, {\it i.e.}, a correlator of a WZW model.
The transition from two dimensions to four is made possible by use of
twistor space $\cp^3$, which is a $\cp^1$-bundle over four-dimensional spacetime.
A conformal field theory is defined in the $\cp^1$ fiber, with spinor momenta
serving as dynamical variables.
Four-dimensional spacetime is then emerged \'a la Penrose \cite{Penrose}.

These ideas had been our understanding of tree-level MHV amplitudes of gluon scattering
until Witten demonstrated further developments in 2003 \cite{Witten:2003nn}.
Remarkably, Witten showed that
Nair's observation is valid even for non-MHV amplitudes in general by introducing
a notion of algebraic curves in twistor space.
This has been carried out partly in connection with string theory.
(For progress in string theory along these lines, see also
\cite{Berkovits:2004hg,Berkovits:2004tx}.)
Soon after then, Cachazo, Svrcek and Witten showed
that these non-MHV amplitudes can also be constructed
in terms of combinations of MHV amplitudes, or vertices, that are connected with
internal off-shell propagators \cite{Cachazo:2004kj}.
This prescription for general amplitudes out of MHV vertices is called the CSW rules.
These rules prove to be of great practical use and replace the conventional
Feynman rules in the computation of helicity-oriented gluon amplitudes.
The CSW rules themselves are therefore a significant result for gauge theories.
For the rigorousness of the CSW rules, one may refer to
\cite{Gukov:2004ei}-\cite{Kiermaier:2008vz}.

Applications of these computations to loop amplitudes have
been under intensive research as well; earlier developments can be found in
\cite{Cachazo:2004zb}-\cite{Brandhuber:2004yw}.
There is also a powerful recursion formula in the calculation of
gluon amplitudes \cite{Britto:2004ap}-\cite{Bern:2005hs}.
For a review of these relatively early developments, see \cite{Cachazo:2005ga}.
An interesting recent discovery along the lines of these developments is
the so-called dual superconformal symmetry in gluon amplitudes \cite{Drummond:2008vq}.
(Dual conformal invariance in planar gluon amplitudes was first indicated in a strong
coupling region \cite{Alday:2007hr}. In the paper \cite{Alday:2007hr},
a computational similarity between gluon amplitudes and Wilson loops
was also observed in a strong coupling region.)
Some of the computational developments in relation with dual superconformal
symmetry can be found in \cite{Brandhuber:2008pf}-\cite{Korchemsky:2009hm}.
For recent reviews of the connection between gluon amplitudes and
Wilson loops in general, one may refer to \cite{Alday:2008yw,Henn:2009bd}.
Since spinor momenta naturally describe massless particles,
it is also appropriate to consider a theory of gravity in the same framework.
There has been much attention to the relation between
Yang-Mills theory and a gravitational theory in this framework.
So far, there have been a number of papers on this subjects; some of them
can be found in \cite{BjerrumBohr:2004wh}-\cite{Drummond:2009ge}.
In the present paper, we shall focus on Yang-Mill theory and leave
the discussion on gravity for an accompanying paper \cite{Abe:hol02}.

The recent computational developments of gluon amplitudes
are indeed significant but the progress are rather technical
with certain emphasis on spinor calculus.
It seems that there is lack of physical interpretation to these
technicalities, that is,
we do not yet fully understand the meaning of a generating functional
or an S-matrix functional for the helicity-oriented gluon amplitudes.
There are in fact many versions of such a functional obtained by
field theoretic analyses \cite{Bianchi:2008pu},\cite{Abe:2004ep}-\cite{Elvang:2008na}.
(For the very recent work, see also \cite{ArkaniHamed:2009si,Mason:2009sa}.)
These results show interesting relations with mathematical properties such as
holomorphicity, integrability and chirality.
However, to some extent, these are obtained inductively form a particular
form of the MHV amplitudes; it is not yet very clear,
at least for the author, whether these functionals can
or cannot be derived from some fundamental principles.
In other words, we would like to obtain the
S-matrix functional of gluon amplitudes
from Nair's original observation somewhat deductively.

Motivated by these considerations, in the present paper, we shall
propose the following novel approach towards the S-matrix functional of gluon amplitudes.
A correlator of a WZW model in general obeys the so-called
Knizhnik-Zamolodchikov (KZ) equation.
There is a very intriguing mathematical fact about the KZ equation, that is,
the monodromy representation of the KZ equation is given by a linear representation
of a braid group.
This is related to the so-called Kohno-Drinfel'd monodromy theorem.
(For details of this theorem, see, {\it e.g.}, \cite{Chari:1994pz}.)
In the spinor-momenta formalism, gauge fields (with some helicities)
are labeled by the numbering index.
Thus, in this formalism, the physical Hilbert space can be defined as
$V_1 \otimes V_2 \otimes \cdots \otimes V_n = V^{\otimes n}$, where $V_ i$
$(i=1,2,\cdots,n)$ denotes a Fock space that creation operators of the $i$-th particle
with helicity $\pm$ act on.
Since gauge fields are bosonic, each $V_i$ is identical
and the Hilbert space is symmetric under permutations of $i$'s.
On the other hand, the physical configuration space $\C$ is defined by
composition of $n$ spinor momenta, preserving the permutation symmetry, {\it i.e.},
$\C = {\bf C}^n / \S_n$, where ${\bf C}$ denotes complex number and $\S_n$ denotes
the rank-$n$ symmetric group.
The fundamental homotopy group of $\C$ then gives the braid group; $\Pi_1 (\C) = \B_n$.
Therefore, the KZ equation or its monodromy representation naturally
arises from the spinor-momenta formalism in twistor space.

The KZ equation has a bialgebraic structure. We take advantage of
this fact to introduce a bialgebraic-valued gauge field which
we call a ``comprehensive'' gauge field $A$.
We construct $A$ to expand an ordinary notion of gauge fields such that
we {\it sum over} the numbering indices labeled on
(ordinary algebra-valued) gauge fields.
The KZ equation can then be interpreted as
an equation expressed as $D \Psi = 0$, where $D$ is
a covariant derivative in terms of $A$,
and $\Psi$ is a function of spinor momenta on $\C$.
(Notice that $\Psi$ can be regarded as a correlation function of a WZW model.)
Further, we can show that the comprehensive gauge field $A$ satisfies
a flatness (or integrability) condition, $D A = 0$.
This arrow us to define a holonomy operator of $A$ in twistor space.
The aim of this paper, along with the accompanying one, is to show that any physical
quantities of gauge theories in twistor space
can universally be generated in terms of such a holonomy operator.
Particularly, in the present paper, we shall obtain a generating functional
that leads to the CSW rules in terms of a variant of the holonomy operator.

The organization of this paper is as follows. In the next section, we
give a brief review of a spinor-momenta formalism in twistor space.
We shall merely review those materials that are necessary for later discussions,
following Nair's lecture note \cite{Nair:2005wh}.
In section 3, we review some known results on the KZ equation, following Kohno's
monograph \cite{Kohno:2002bk}.
From these results, we see that it is natural to introduce the above mentioned
``comprehensive'' gauge field.
For this purpose, we find that it is suitable
to introduce a bialgebraic property for gauge fields.
Holonomy operators for the comprehensive gauge fields are also defined in this section.
In section 4, in addition to the bialgebraic structure, we introduce other physical
properties to the gauge fields in twistor space so that helicity information
of the particles is appropriately incorporated.
For this purpose, we impose $\N = 4$ extended supersymmetry on the holonomy
operator. We show that the supersymmetric
holonomy operator naturally leads to a generating functional for the
MHV amplitudes.
In section 5, we generalize this approach to non-MHV amplitudes.
In a language of functional derivative, this means an introduction
of a contraction operator to the MHV generating functional.
The CSW rules can be realized by such an operator in this context.
We find that an S-matrix functional for non-MHV amplitudes
can be expressed in terms of the contraction operator and the
supersymmetric holonomy operator.
Lastly, we shall present some concluding remarks.

\section{Spinor momenta, twistor space and Nair measure}

In this section, we review a spinor-momenta formalism
in twistor space, following Nair's lecture note \cite{Nair:2005wh}.

\noindent
\underline{Spinor momenta}

Massless particles, such as gluons and gravitons, have
momenta $p_\mu$ $(\mu = 0,1,2,3)$ which obey the on-shell condition $p^2 =
\eta^{\mu\nu} p_\mu p_\nu = p_{0}^{2} - p_{1}^{2} - p_{2}^{2} - p_{3}^{2} =0$,
where $\eta^{\mu\nu} = (+,-,-,-)$ denotes the Minkowski metric.
From this condition, it is possible to express the four-momentum $p_\mu$ as
a $(2 \times 2)$-matrix
\beq
p^{A}_{\, \Ad}  = (\si^\mu)^{A}_{\, \Ad} \, p_\mu   =
        \left(
            \begin{array}{cc}
              p_0 + p_3 & p_1 - i p_2 \\
              p_1 + i p_2 & p_0 - p_3 \\
            \end{array}
         \right) \equiv  u^A \bu_\Ad
\label{2-1}
\eeq
where both $A$ and $\Ad$ take values of $(1,2)$. $\si^\mu$ is given by
$\si^\mu = ( {\bf 1}, \si^i)$, where $\si^i$ ($i = 1,2,3$) and ${\bf 1}$
are the Pauli matrices and the $(2 \times 2)$ identity matrix, respectively.
In the above equation, $u^A$ and $\bu_\Ad$ denote two-component spinors.
An explicit choice for these is given by
\beq
u^A = {1 \over \sqrt{p_0 - p_3}} \left(
        \begin{array}{c}
          {p_1 - i p_2} \\
          {p_0 - p_3} \\
        \end{array}
      \right) \, , ~~
\bu_\Ad  = {1 \over \sqrt{p_0 - p_3}}
    \left(
         \begin{array}{c}
           {p_1 + i p_2 } \\
           {p_0 - p_3} \\
         \end{array}
    \right)
\label{2-2}
\eeq
where we follow a convention to express a spinor as a column vector.
Requiring that  $p_\mu$ be real, we can take $\bu_\Ad$ as a
conjugate of $u^A$, $\bu_\Ad = (u^A)^*$.
From (\ref{2-1}), we see that the four-momentum is invariant under
\beq
u^A \rightarrow e^{i \phi} u^A \, , ~~~~~ \bu_\Ad \rightarrow e^{-i \phi} \bu_\Ad
\label{2-3}
\eeq
where $\phi$ represents a $U(1)$ phase parameter.
Thus there is a phase ambiguity in the definition of $u^A$ and $\bu_\Ad$.

From these parametrizations, we find that four-momenta of massless particles
can be described by the spinor $u^A$. This is called the spinor momenta.
Lorentz transformations of $u^A$ are given by
\beq
u^A \rightarrow (g u)^A
\label{2-4}
\eeq
where $g \in SL(2, {\bf C})$
is a $(2 \times 2)$-matrix representation of $SL(2,{\bf C})$;
the complex conjugate of this relation leads to Lorentz transformations of $\bu_\Ad$.
Four-dimensional Lorentz transformations are realized by a
combination of these, {\it i.e.}, the four-dimensional Lorentz symmetry is
given by $SL(2,{\bf C}) \times SL(2,{\bf C})$.
Scalar products of $u^A$'s or $\bu_\Ad$'s, which are invariant under the
corresponding $SL(2,{\bf C})$, are expressed as
\beq
 u_i \cdot u_j \equiv (u_i u_j) =   \ep_{AB} u_{i}^{A}u_{j}^{B} \, , ~~~~~
 \bu_i \cdot \bu_j \equiv [\bu_i \bu_j]  = \ep^{\Ad \Bd} \bu_{i \, \Ad}
 \bu_{j \, \Bd}
\label{2-5}
\eeq
where $\ep_{AB}$ is the rank-2 Levi-Civita tensor.
This can be used to raise or lower the indices, {\it e.g.}, $u_B = \ep_{AB}u^A$.
Notice that these products are zero when $i$ and $j$ are identical.
In what follows, we can assume $1 \le i < j \le n$ without loss of generality.

In the present paper, we are interested in a theory with conformal invariance.
More precisely, our primal interest is $\N = 4$ super Yang-Mills theory at the tree level.
So we shall now require scale invariance to
the spinor momentum
\beq
u^A \sim \la u^A \, , ~~~~~ \la \in {\bf C} - \{ 0 \}
\label{2-6}
\eeq
where $\la$ is non-zero complex number.
With this identification, we can regard
the spinor momentum $u^A$ as a homogeneous coordinate
of the complex projective space $\cp^1$.
The local coordinate of $\cp^1$ is represented by
a single complex variable $z$. This can be related to
$u^A$ by the following parametrization.
\beq
u^A = \frac{1}{\sqrt{p_0 - p_3}} \left(
        \begin{array}{c}
          {p_1 - i p_2} \\
          {p_0 - p_3} \\
        \end{array}
      \right)
\equiv \left(
        \begin{array}{c}
          \al \\
          \bt \\
        \end{array}
      \right)
= \al \left(
        \begin{array}{c}
          1 \\
          z \\
        \end{array}
      \right)
,~~~~ z = {\bt \over \al} ~~ (\al \ne 0)
\label{2-7}
\eeq
The local complex coordinate of
$\cp^1$ can be taken as $z = \bt / \al$ except in the vicinity of $\al = 0$,
where we can instead use $\al / \bt$ as the local coordinate.
In what follows, we shall consider $n$ distinct spinor momenta
$u^{A}_{i}$ ($i = 1,2, \cdots, n$) which correspond to the complex
coordinates $z_i$ with an assumption of $\al \ne 0$ for each of $z_i$.

One of the advantages to use spinor momenta comes from the fact that
certain physical quantities, {\it i.e.},
the maximal helicity violating (MHV) amplitudes for gluons,
can be expressed entirely in terms of the scalar product of spinor momenta.
Further, recent developments in the understanding of gluon amplitudes in general
support a case that this scalar product serves as a unit of
any physical quantities or observables.
This idea leads to the following covariant form as a unit of
observables
\beq
\om_{ij} \equiv d \log(u_i u_j) = \frac{d(u_i u_j)}{(u_i u_j)}
\label{2-8}
\eeq
where we introduce a logarithmic quantity so that the scale invariance
is encoded automatically.
One may think of this
as a {\it propagator unit} since a propagator on $\cp^1$ is essentially
given by $\log (u_i u_j)$.
Notice that $(u_i u_j)$ can be written as $(u_i u_j) = - \al_i \al_j (z_i - z_j)$
and that $\al$'s are functions of $(z, \bz)$ in general.
Because of the scale invariance of $\cp^1$, however, we can fix these $\al$'s and
define $(z, \bz)$ as variations from the fixed values. Thus the above one-form
can also be written as $\om_{ij} = d \log(z_i - z_j)$.
In the next section, we shall use this form as a basic quantity.

\noindent
\underline{Twistor space}

Twistor space is defined by a four-component spinor
$Z_I =( \pi^A, v_\Ad)$ $(I = 1,2,3,4)$ where $\pi_A$ and $v_\Ad$ are two-component
complex spinors.
From this definition, it is easily understood that
twistor space is represented by $\cp^3$; $Z_I$ can be interpreted as
the complex homogeneous coordinates of
$\cp^3$, satisfying the following relation.
\beq
Z_I  \sim \la Z_I \, , ~~~~~ \la \in {\bf C} - \{ 0 \}
\label{2-9}
\eeq
In twistor space, the relation between $\pi^A$ and $v_\Ad$ is defined as
\beq
v_\Ad \, = \, x_{\Ad A} \pi^A
\label{2-10}
\eeq
With this condition, the relation (\ref{2-9}) is realized by the scale invariance
of $\pi^A$, as shown in (\ref{2-6}) for $u^A$.
In (\ref{2-10}), $x_{\Ad A}$ represent local coordinates on $S^4$;
this can be understood from the fact that $\cp^3$ is a $\cp^1$-bundle over $S^4$.
We consider that the $S^4$ describes a four-dimensional compact spacetime.
A flat spacetime may be obtained by considering a neighborhood of this $S^4$.

In terms of $x_{\Ad A}$, the conventional four-dimensional coordinates $x_\mu$
can be expressed as
\beq
x_{\Ad A} = x_\mu (\si^\mu)_{A \, \Ad}  =
        \left(
            \begin{array}{cc}
              x_0 + x_3 & x_1 - i x_2 \\
              x_1 + i x_2 & x_0 - x_3 \\
            \end{array}
         \right)
\label{2-11}
\eeq
Owing to the scale invariance of $\pi^A$ along with the twistor-space
condition (\ref{2-10}), we find that the
size or radius of $S^4$ is irrelevant in the present context.
Notice also that in twistor space the spacetime coordinates $x_{\Ad A}$
are emergent quantities.
Four-dimensional diffeomorphism is therefore realized by $u^A \rightarrow u'^A$,
rather than $x_{\Ad A} \rightarrow {x'}_{\Ad A}$.

{\it
What is essential in the spinor-momenta formalism in twistor space is to
identify a $\cp^1$ fiber of twistor space with a $\cp^1$ on which the spinor momenta
are defined. In other words, we identify $\pi_A$ with the spinor momenta $u_A$
so that we can essentially describe four-dimensional physics
in terms of the coordinates of $\cp^1$.
}

The twistor-space condition (\ref{2-10}) now becomes $v_\Ad = x_{\Ad A} u^A$ and
using this relation we can easily identify the product $v_\Ad \bu^\Ad$
as $x_{\Ad A} p^{A \Ad}$.
Using the rule of scalar products (\ref{2-5}), we can express $v_\Ad$ as
$v_{\Ad} = x_{\Ad 1} u^2 - x_{\Ad 2} u^1$.
Thus, in terms of the Minkowski coordinates, the product $x_{\Ad A} p^{A \Ad}$
can be calculated as
\beq
x_{\Ad A} p^{A \Ad} = 2 (x_0 p_0 - x_1 p_1 - x_2 p_2 - x_3 p_3 ) = 2 x_{\mu} p^{\mu}
\label{2-12}
\eeq

In the spinor-momenta formalism, twistor variables are represented by
$Z_I =(u^A, v_\Ad)^{T}$.
Linear transformations of $Z_I$ are realized by
\beq
Z_I  \rightarrow (g Z)_I = g_{I}^{\, J} Z_J
\label{2-13}
\eeq
Here $g \in SU(2,2)$ denotes the $(4 \times 4)$-matrix representation
of $SU(2,2)$.
Thus the above transformations correspond
conformal transformations of twistor variables.
Generators of these transformations can be classified as follows.
\beqar
\nonumber
J_{AB} = \hf \left( u_A \frac{\d}{\d u^B} + u_B \frac{\d}{\d u^A} \right)
& : & \mbox{holomorphic Lorentz transformation} \\
\nonumber
J_{\Ad \Bd} = \hf \left( v_\Ad \frac{\d}{\d  v^\Bd} +
v_\Bd \frac{\d}{\d v^\Ad} \right)
& : & \mbox{antiholomorphic Lorentz transformation} \\
P^{A \Ad} = u^A {\d \over {\d v_\Ad}}
& : & \mbox{spacetime translation}
\label{2-14}\\
\nonumber
K_{\Ad A} = v_\Ad {\d \over {\d u^A}}
& : & \mbox{special conformal transformation} \\
\nonumber
D = v_\Ad \frac{\d}{\d v_\Ad} - u^A \frac{\d}{\d u^A}
& : & \mbox{dilatation}
\eeqar

Comparing (\ref{2-14}) with (\ref{2-1}), we can easily see that $\bu^\Ad$ corresponds to
$\frac{\d}{\d v_\Ad}$. This suggests that we can relate a function of $(u, \bu)$ with
a function of $(u, v)$ by a Fourier transform integral
\beq
f(u, v) = \frac{1}{4}
\int \frac{d^2 \bu}{(2 \pi )^2} \, f(u, \bu) \, e^{\frac{i}{2} v_\Ad \bu^\Ad}
\label{2-15}
\eeq
Notice that we choose the exponent so that it becomes
the four-dimensional product $i x_{\mu}p^{\mu}$ by use of (\ref{2-12}).
As we will discuss later, this
is an illuminating fact for extraction of four-dimensional quantities out of
functions of $(u, \bu)$.
The above integral (\ref{2-15}) is
referred to as a Fourier transform in twistor space.

In (\ref{2-14}), there are two types of Lorentz generators.
These can be expressed as
\beq
J_{AB} = \hf \left( x_{A \Ad} \, p^{\Ad}_{B} + x_{B \Bd} \, p^{\Bd}_{A} \right)
\, , ~~
J_{\Ad \Bd} = \hf \left( x_{\Ad A} \, p^{A}_{\Bd} + x_{\Bd B} \, p^{B}_{\Ad} \right)
\label{2-16}
\eeq
Thus these are indeed corresponding to four-dimensional Lorentz generators.
That the four-dimensional generators split into $J_{AB}$ and $J_{\Ad \Bd}$
in twistor space comes from a basic fact of the spinor-momenta formalism,
{\it i.e.} we have two types of contraction of indices,
involving those with dots or those without dots, as seen in the above expression.

A helicity of a massless particle can be determined by
a Pauli-Lubanski spin vector $S^\mu$. Conventionally, this is defined as
$S^\mu = \ep^{\mu \nu \al \bt} p_\nu J_{\al \bt}$, where $J_{\al \bt}$ denotes
a four-dimensional Lorentz generator.
For massless particles $(p^2 = 0)$, this can be characterized by $S^2 = 0$ and
$S \cdot p =0$ so that we can define a helicity $h$ as $S^\mu = h \, p^{\mu}$.
We now define an analog of the Pauli-Lubanski spin vector in twistor space as
\beq
S_{\Bd B} = p^{A}_{\Bd} {\cal J}_{AB} =p^{A}_{\Bd} \left( \del_{AB} - J_{AB} \right)
= p_{\Bd B} \left( 1 - \hf u^A \frac{\d}{\d u^A} \right)
\label{2-17}
\eeq
where we introduce an expression ${\cal J}_{AB} = \del_{AB} - J_{AB}$ so
that we can relate $S_{\Bd B}$ with the four-dimensional $S^\mu$.
Notice that this is not the only choice of the spin vector.
In fact, we can use any ${\cal J}_{AB}$ that is written as
${\cal J}_{AB} = c \, \del_{AB} - J_{AB}$, with $c$ being any
constant, since this ${\cal J}_{AB}$ always leads to the relation
$S_{\Ad A}p^{A \Ad} = S^2 = 0$.
A natural choice would be $c=0$, which has been taken in
\cite{Nair:1988bq,Abe:2004ep}.
In the present paper, we rather choose $c=1$ for later convenience.

The helicity operator is thus given by
\beq
h = 1 -  \hf u^A \frac{\d}{\d u^A}
\label{2-18}
\eeq
This shows that the helicity of the $i$-th gluon
is essentially given by the degree of homogeneity in $u_i$.

\noindent
\underline{Nair measure}

From our specific parametrization for the spinor momentum in (\ref{2-2}),
we can do calculate a Lorentz invariant measure in terms of $(u, \bu)$ as
\beqar
\nonumber
d \mu (p) \equiv
\frac{d^3 p}{(2 \pi)^3} \frac{1}{2 p_0}
&=&
\frac{1}{(2 \pi)^3} \frac{({\bar \al} \al) d ({\bar \al} \al)}{2} \frac{ dz d \bz}{(-2i)}
\\
\label{2-19}
&=&
\frac{1}{4} \left[
\frac{u \cdot du}{2 \pi i} \frac{d^2 \bu}{(2 \pi)^2} -
\frac{\bu \cdot d \bu}{2 \pi i} \frac{d^2 u}{(2 \pi)^2}
\right]
\eeqar
This is the so-called Nair measure.
Since the spinor-momenta formalism is manifestly Lorentz invariant,
the Nair measure is what we should use to derive any physical observables.
Notice that an integration over $\frac{u \cdot du}{2 \pi i}$
corresponds to a contour integral on $\cp^1$.
From (\ref{2-7}), the measure of the contour integral can be calculated as
\beq
\oint \frac{(u_i d u_i)}{2 \pi i} = \al^2 \oint \frac{d z_i}{2 \pi i}
\label{2-20}
\eeq
This integral corresponds to the so-called Penrose integral in twistor space.
If integrands are antiholomorphic, there are no singularities in the $z$-space and
this part of the Nair measure vanishes.
Likewise, if integrands are holomorphic, the second term in (\ref{2-19}) vanishes.

\section{Braid group, KZ equation and holonomies}

As we have briefly mentioned in the introduction, we take advantage of the knowledge of
mathematical results on the Knizhnik-Zamolodchikov (KZ) equation.
In this section we review these results, following
Kohno's textbook \cite{Kohno:2002bk}.
We also consider applications of these results to the spinor-momenta formalism.

\noindent
\underline{Braid group}

The Hilbert space of the spinor-momenta formalism
for the description of gluons is given by
$V^{\otimes n} = V_1 \otimes V_2 \otimes \cdots \otimes V_n$
where $V_ i$ $(i=1,2,\cdots,n)$ denotes a Fock space that creation operators
of the $i$-th particle with helicity $\pm$ act on.
Such operators can be expressed as $a_{i}^{(\pm)}$, with $(\pm)$
denoting helicities of the gluons.
Notice that $a_{i}^{(-)}$ can be given by the conjugate of $a_{i}^{(+)}$,
$a_{i}^{(-)} = (a_{i}^{(+)})^*$, and vice versa.
These can be interpreted as
ladder operators which form a part of the $SL(2, {\bf C})$ algebra.
The algebra can be expressed as
\beq
[ a_{i}^{(+)}, a_{j}^{(-)}] = 2 a_{i}^{(0)} \, \del_{ij}  \, , ~~~
[ a_{i}^{(0)}, a_{j}^{(+)}] = a_{i}^{(+)} \, \del_{ij} \, , ~~~
[ a_{i}^{(0)}, a_{j}^{(-)}] = - a_{i}^{(-)} \, \del_{ij}
\label{3-1}
\eeq
where Kronecker's deltas show that
the non-zero commutators are obtained only when $i = j$.
The remaining of commutators, those expressed otherwise, all vanish.

A spinor momentum is defined on $\cp^1$ and its local coordinate can be
expressed by the complex variable $z$.
Since there are $n$ momenta,
the physical configuration space is given by $\C = {\bf C}^n / \S_n$,
where $\S_n$ is the symmetric group of rank $n$.
The $\S_n$ arises from the fact that gluons are bosons.
It is well known that the fundamental homotopy group of $\C = {\bf C}^n / \S_n$
is given by the braid group, $\Pi_1 (\C) = \B_n$.
The braid group $\B_n$ has generators, $b_1 , b_2 , \cdots ,
b_{n-1}$, and they satisfy the following relations.
\beqar
\nonumber
b_i b_{i+1} b_i &=& b_{i+1} b_i b_{i+1} \, ,~~~ \mbox{if}~ |i - j| = 1 \\
b_i b_j &=& b_j b_i \, , ~~~~~~~~~~~ \mbox{if}~ |i - j| > 1
\label{3-2}
\eeqar
where we identify $b_n$ with $b_1$.
To be mathematically rigorous, the ${\bf C}$ of $\C = {\bf C}^n / \S_n$ should
be replaced by $\cp^1$ which is represented by the local coordinate $z$.
Since ${\bf C}$ can be obtained from $\cp^1$ by excluding points at infinity,
{\it i.e.}, ${\bf C} = \cp^1 / \{ \infty \}$, the replacement
can be done with ease.
At the level of braid generators, this can be
carried out by imposing the following relation \cite{Chari:1994pz}
\beq
(b_1 b_2 \cdots b_{n-2} b_{n-1} ) (b_{n-1} b_{n-2} \cdots b_2 b_1) = 1
\label{3-3}
\eeq
The braid group that satisfies this condition on top of (\ref{3-2}) is
called a {\it sphere} braid group $\B_n (\cp^1)$, while the previous one is called
a {\it pure} braid group $\B_n ({\bf C}) = \B_n $.
Thus, bearing in mind this subsidiary condition, we can identify $\C = {\bf C}^n / \S_n$
as the physical configuration space of interest.

\noindent
\underline{KZ equation and emergence of bialgebra}

We now consider logarithmic differential one-forms
\beq
\om_{ij} = d \log (z_i - z_j) = \frac{ d z_i - d z_j}{z_i - z_j}
\label{3-4}
\eeq
These satisfy the identity
\beq
\om_{ij} \wedge \om_{jk} + \om_{jk} \wedge \om_{ik} + \om_{ik} \wedge \om_{ij} = 0
\label{3-5}
\eeq
where the indices are ordered as $i < j < k$.

The Knizhnik-Zamolodchikov (KZ) equation
is an equation that a correlator of a WZW model, or more generally a
function of the physical configuration $\C$, satisfies.
Such a function can be evaluated as a vacuum expectation value of
operators acting on the Hilbert space $V^{\otimes n}$.
Let us denote such a function as $\Psi (z_1, \cdots , z_n)$. Then the
KZ equation is defined as
\beq
\frac{\d \Psi }{ \d z_i} =  \frac{1}{\kappa}
\sum_{~ j \, (j \ne i)} \frac{\Om_{ij} \Psi}{z_i - z_j}
\label{3-6}
\eeq
where $\kappa$ is a non-zero constant and $\Om_{ij}$ can be expressed as
\beq
\Om_{ij} = a_{i}^{(+)} \otimes a_{j}^{(-)} + a_{i}^{(-)} \otimes a_{j}^{(+)}
+ 2 a_{i}^{(0)} \otimes a_{j}^{(0)}
\label{3-7}
\eeq
Notice that $\Om_{ij}$ is a bialgebraic operator and that its action on
$V^{\otimes n} = V_1 \otimes V_2 \otimes \cdots \otimes V_n$ can be written as
\beq
\sum_{\mu} 1 \otimes \cdots \otimes 1 \otimes \rho_i (I_{\mu})
\otimes 1 \otimes \cdots \otimes 1 \otimes \rho_j (I_{\mu}) \otimes 1 \otimes \cdots
\otimes 1
\label{3-8}
\eeq
where $I_\mu$ ($\mu = 0,1,2$) are elements of the $SL(2, {\bf C})$ algebra
and $\rho$ denotes its representation.
Introducing the following one-form
\beq
\Om =  \frac{1}{\kappa} \sum_{1 \le i < j \le n} \Om_{ij} \, \om_{ij} \, ,
\label{3-9}
\eeq
we can rewrite the KZ equation (\ref{3-6}) as a total differential equation
\beq
D \Psi = (d - \Om) \Psi = 0
\label{3-10}
\eeq
where $D = d - \Om$ can be regarded as a covariant exterior derivative.
$\Om$ is called the KZ connection in mathematics.

What is remarkable for physicists is that one can prove
the flatness of the KZ connection, {\it i.e.}, $d \Om - \Om \wedge \Om
= 0$, just by using the relations on $\Om_{ij}$ given by \cite{Kohno:2002bk}
\beqar
\label{3-11}
[ \Om_{ij} , \Om_{kl} ] &=& 0  ~~~~~ \mbox{($i,j,k,l$ are distinct)} \\  \label{3-12}
[ \Om_{ij} + \Om_{jk} , \Om_{ik} ] &=& 0  ~~~~~ \mbox{($i,j,k$ are distinct)}
\eeqar
From (\ref{3-7}), we find that these are commutators of bialgebraic operators, which
can be evaluated in terms of the commutators among $a_{i}^{(\pm)}$ and $a_{i}^{(0)}$.
The relation (\ref{3-11}) is then obvious, since $i,j,k,l$ are distinct.
It is instructive to show the relation (\ref{3-12}). So we sketch the proof in the following.
Non-vanishing terms of $[ \Om_{ij} , \Om_{ik} ]$ can be expanded as
\beqar
\nonumber
[ \Om_{ij} , \Om_{ik} ] &=&
[a_{i}^{(+)} \otimes a_{j}^{(-)} , a_{i}^{(-)} \otimes a_{k}^{(+)} ] +
[a_{i}^{(+)} \otimes a_{j}^{(-)} , 2 a_{i}^{(0)} \otimes a_{k}^{(0)} ] \\ \nonumber
&& \!\!\!\!\! + \, [a_{i}^{(-)} \otimes a_{j}^{(+)} , a_{i}^{(+)} \otimes a_{k}^{(-)} ]+
[a_{i}^{(-)} \otimes a_{j}^{(+)} , 2 a_{i}^{(0)} \otimes a_{k}^{(0)} ] \\
&& \!\!\!\!\! + \, [2 a_{i}^{(0)} \otimes a_{j}^{(0)} , a_{i}^{(+)} \otimes a_{k}^{(-)} ] +
[2 a_{i}^{(0)} \otimes a_{j}^{(0)} , a_{i}^{(-)} \otimes a_{k}^{(+)} ]
\label{3-13}
\eeqar
The first two terms can be evaluated as
\beqar
\nonumber
&& [ a_{i}^{(+)} , a_{i}^{(-)} ] \otimes a_{j}^{(-)} \otimes a_{k}^{(+)} +
[ a_{i}^{(+)}, 2 a_{i}^{(0)} ] \otimes a_{j}^{(-)} \otimes a_{k}^{(0)} \\ \nonumber
&=&
2 a_{i}^{(0)} \otimes a_{j}^{(-)} \otimes a_{k}^{(+)} -
2 a_{i}^{(+)} \otimes a_{j}^{(-)} \otimes a_{k}^{(0)}
\eeqar
where we use the following definition of commutators for bialgebraic operators.
\beqar
\nonumber
[ a_i \otimes b_i , a_j \otimes b_j ] &=& [a_i , a_j ] \otimes b_i \otimes b_j \\ \nonumber
&& \!\!\!\! + \, a_i \otimes [b_i , a_j ] \otimes b_j \\
&& \!\!\!\! + \, a_i \otimes a_j \otimes [b_i , b_j]
\label{3-14}
\eeqar
Here $a_i$ and $b_i$ $(i = 1,2,\cdots , n)$ denote arbitrary (algebraic) operators
in a usual sense.
Similarly, the rest of the terms in (\ref{3-13}) can be evaluated and we obtain
\beqar
\nonumber
[ \Om_{ij} , \Om_{ik} ] &=& 2 \left( \,
a_{i}^{(0)} \otimes a_{j}^{(-)} \otimes a_{k}^{(+)} -
a_{i}^{(+)} \otimes a_{j}^{(-)} \otimes a_{k}^{(0)} \right. \\ \nonumber
&& \! - \, a_{i}^{(0)} \otimes a_{j}^{(+)} \otimes a_{k}^{(-)} +
a_{i}^{(-)} \otimes a_{j}^{(+)} \otimes a_{k}^{(0)} \\
&& \left. \! + \, a_{i}^{(+)} \otimes a_{j}^{(0)} \otimes a_{k}^{(-)}
- \, a_{i}^{(-)} \otimes a_{j}^{(0)} \otimes a_{k}^{(+)} \, \right)
\label{3-15}
\eeqar
We can also work out for $[ \Om_{jk} , \Om_{ik} ]$ and easily find that
$[ \Om_{jk} , \Om_{ik} ] = - [ \Om_{ij} , \Om_{ik} ]$.
Thus, the relation (\ref{3-12}) indeed holds.
Since $d \Om = 0$, we need to show $\Om \wedge \Om = 0$ for the flatness of $\Om$.
This can be confirmed by use of the identity (\ref{3-5})
and the relations (\ref{3-11}), (\ref{3-12}).

We now consider applications of these mathematical results to
the spinor-momenta formalism for gluons.
As mentioned before, the physical operators of gluons are given by $a_{i}^{(\pm)}$.
The operator $\Om_{ij}$ may not be appropriate to describe gluons since
its action on the Hilbert space is represented by (\ref{3-8}), which involves $a_{i}^{(0)}$.
We need to modify $\Om_{ij}$'s so that the operators $a_{i}^{(0)}$
are treated somewhat unphysically.
We then introduce a ``comprehensive'' gauge one-form
\beq
A =  \frac{1}{\kappa} \sum_{1 \le i < j \le n} A_{ij} \, \om_{ij}
\label{3-16}
\eeq
where $A_{ij}$ is defined as a bialgebraic operator
\beq
A_{ij}
= a_{i}^{(+)} \otimes a_{j}^{(0)} + a_{i}^{(-)} \otimes a_{j}^{(0)}
\label{3-17}
\eeq
Notice that $A_{ij}$ also satisfies the relations (\ref{3-11}), (\ref{3-12}).
It is sufficient to show $[A_{ij} + A_{jk} , A_{ik}] = 0$ here and
this can be checked as follows.
\beqar
\nonumber
[A_{ij} , A_{ik} ] &=& [a_{i}^{(+)} \otimes a_{j}^{(0)} , a_{i}^{(-)} \otimes a_{k}^{(0)} ]
+ [a_{i}^{(-)} \otimes a_{j}^{(0)} , a_{i}^{(+)} \otimes a_{k}^{(0)} ] \\
&=& 2 a_{i}^{(0)} \otimes a_{j}^{(0)} \otimes a_{k}^{(0)}
 \, - \, 2 a_{i}^{(0)} \otimes a_{j}^{(0)} \otimes a_{k}^{(0)}
 \, = \, 0
\label{3-18}
\eeqar
We can also find that $[A_{jk}, A_{ik}]$ vanishes with ease since all terms of this
commutator are proportional to $[a_{k}^{(0)}, a_{k}^{(0)}]$ which is zero by definition.

Since the relations (\ref{3-11}) and (\ref{3-12}) are the only conditions
for the flatness of the connection, from the above results we find
\beq
DA = dA - A \wedge A = - A \wedge A = 0
\label{3-19}
\eeq
where we redefine $D$ as a covariant exterior derivative $D = d - A$.
The relations (\ref{3-11}) and (\ref{3-12}) are crucial for the flatness
of the comprehensive gauge field.
These relations are called
infinitesimal braid relations in mathematical literature.

\noindent
\underline{Holonomy operators}

The KZ equation corresponding to $A$ is then
expressed as $D \Psi = (d - A) \Psi =0$.
This shows that any function $\Psi$ on $\C$ is covariantly constant, which
means that the function $\Psi$ is parallel along an open path on $\C$.
The comprehensive connection one-form $A$ satisfies the flatness (or integrability)
condition (\ref{3-19}).
Mathematical circumstances are then precisely the same as those in
the case for the construction of Wilson loop operators.
Thus it is mathematically possible to define holonomy operators of the
comprehensive gauge one-form $A$.
The particular form of $A$ in (\ref{3-16}), however,
suggests that a single use of $A$
is not appropriate for the construction of holonomies because
the use of differential forms indicates that the physically relevant configuration
is represented by a differential $n$-manifold and
an integration over this manifold requires a differential $n$-form for its
integrand.
We can then define a holonomy operator of $A$ as
\beq
\Theta_{R, \ga} \, = \, \Tr_{R, \ga} \, \Path \exp \left[
\sum_{m=0}^{\infty} \oint_{\ga} \underbrace{A \wedge A \wedge \cdots \wedge A}_{m}
\right]
\label{3-20}
\eeq
where $\ga$ denotes a closed path on $\C$ along which the integral is evaluated and
$R$ denotes the representation of the gauge group as usual.
The color degree of freedom can be attached to the physical operator $a_{i}^{(\pm)}$
in (\ref{3-17}) as
\beq
a_{i}^{(\pm)} = t^{c_i} \, a_{i}^{(\pm)c_i}
\label{3-21}
\eeq
where $t^{c_i}$'s are the generators of the gauge group in the $R$-representation.
In the present paper, we consider $SU(N)$ gauge groups.
The symbol $\Path$ denotes a ``path ordering'' of the numbering indices.
The meaning of this symbol can be understood as follows.
The exponent in (\ref{3-20}) can be expanded as
\beq
\sum_{m=0}^{\infty} \oint_{\ga} \underbrace{A \wedge \cdots \wedge A}_{m}
= \sum_{m=0}^{\infty} \oint_{\ga}  \sum_{ (i<j) } A_{i_1 j_1} A_{i_2 j_2} \cdots A_{i_m j_m}
\bigwedge_{k=1}^{m} \om_{i_k j_k}
\label{3-22}
\eeq
where $(i < j)$ means that the set of indices $(i_1, j_1, \cdots , i_m , j_m)$
are ordered such that $1 \le i_1 < j_1 \le m$,$\cdots$,$1 \le i_m < j_m \le m$.
Notice that, for $m \le 1$, the exponent is defined as zero.
The symbol $\Path$ means that these indices are further received ordering conditions
$1 \le i_1 < i_2 < \cdots < i_m \le m$ and $2 \le j_1 < j_2 < \cdots < j_m \le m+1$
($m+1$ is to be identified with $1$). This automatically leads to
the following expression
\beq
\Path \sum_{m=0}^{\infty} \oint_{\ga} \underbrace{A \wedge \cdots \wedge A}_{m}
= \sum_{m=0}^{\infty} \oint_{\ga}  A_{1 2} A_{2 3} \cdots A_{m 1}
\, \om_{12} \wedge \om_{23} \wedge \cdots \wedge \om_{m 1}
\label{3-23}
\eeq
which shows that the basis of
the holonomy operators (\ref{3-20}) is given by $\om_{i \, i+1}$
under the ``path ordering'' operation for the numbering indices.

Now the commutator $[A_{12}, A_{23}]$ can be calculated as
\beqar
\nonumber
[A_{12}, A_{23}]
&=& a_{1}^{(+)} \otimes a_{2}^{(+)} \otimes a_{3}^{(0)}
- a_{1}^{(+)} \otimes a_{2}^{(-)} \otimes a_{3}^{(0)}  \\
&& \!\!\! + \, a_{1}^{(-)} \otimes a_{2}^{(+)} \otimes a_{3}^{(0)}
- a_{1}^{(-)} \otimes a_{2}^{(-)} \otimes a_{3}^{(0)}
\label{3-24}
\eeqar
Equation (\ref{3-23}) is then written as
\beqar
\nonumber
\Path \sum_{m=0}^{\infty} \oint_{\ga} \underbrace{A \wedge \cdots \wedge A}_{m}
&=& \sum_{m=0}^{\infty} \oint_{\ga}  A_{1 2} A_{2 3} \cdots A_{m 1}
\, \om_{12} \wedge \om_{23} \wedge \cdots \wedge \om_{m 1} \\
\nonumber
&=& \sum_{m=0}^{\infty} \frac{1}{2^{m+1}} \sum_{(h_1, h_2, \cdots , h_m)}
(-1)^{h_1 + h_2 + \cdots + h_m} \\
&& ~~~ \times \,
a_{1}^{(h_1)} \otimes a_{2}^{(h_2)} \otimes \cdots \otimes a_{m}^{(h_m)}
\, \oint_{\ga} \om_{12} \wedge \cdots \wedge \om_{m1}
\label{3-25}
\eeqar
where $h_{i} = \pm = \pm 1$ ($i=1,2,\cdots, m$) denotes
the helicity of the $i$-th particle.
In the above expression, we define
$a_{1}^{(\pm)} \otimes a_{2}^{(h_2)} \otimes
\cdots \otimes a_{m}^{(h_m)} \otimes a_{1}^{(0)}$ as
\beqar
\nonumber
a_{1}^{(\pm)} \otimes a_{2}^{(h_2)} \otimes \cdots \otimes a_{m}^{(h_m)} \otimes a_{1}^{(0)}
& \equiv &
\hf [a_{1}^{(0)} , a_{1}^{(\pm)}] \otimes a_{2}^{(h_2)} \otimes \cdots \otimes a_{m}^{(h_m)} \\
&=&
\pm \hf a_{1}^{(\pm)} \otimes a_{2}^{(h_2)} \otimes \cdots \otimes a_{m}^{(h_m)}
\label{3-26}
\eeqar
where we implicitly use an antisymmetric property for the indices $(1,2, \cdots , m)$
as indicated in (\ref{3-22}) or (\ref{3-23}).

The holonomy operator (\ref{3-20}) is a general solution to the KZ equation
$D \Psi = (d - A) \Psi = 0$ and is mathematically well-defined.
Since $A$ satisfies the flat condition (\ref{3-19}), a linear transformation
of the holonomy operator depends only the homotopy class of the
loop $\ga$ on $\C = {\bf C}^n / \S_n$.
Therefore, the holonomy operator $\Theta_{R, \ga}$ gives
a linear representation of the braid group $\B_n$.
This representation is called the monodromy representation of the KZ equation.

The information about the braid group is in one-to-one correspondence with
permutation of the numbering indices. Namely, a distinct permutation can be
represented by a distinct product of braid generators.
A ``trace'' over such a product is then encoded by a sum over possible
permutations. We symbolize this notion by $\Tr_{\ga}$ which can be expressed
as
\beq
\Tr_{\ga} \Path \sum_{m = 0}^{\infty} \oint_{\ga}
\underbrace{A \wedge \cdots \wedge A}_{m}
= \sum_{m=0}^{\infty}
\sum_{\si \in \S_{m-1}} \oint_{\ga}  A_{1 \si_2} A_{\si_2 \si_3} \cdots A_{\si_m 1}
\, \om_{1 \si_2} \wedge \om_{\si_2 \si_3} \wedge \cdots \wedge \om_{\si_m 1}
\label{3-27}
\eeq
where the summation of $\S_{m-1}$ is taken over the
permutations of the elements $\{2,3, \cdots, m \}$, {\it i.e.},
the relevant permutations are given by
$\si=\left(%
\begin{array}{c}
  2 \, 3 \cdots m \\
  \si_2 \si_3 \cdots \si_m \\
\end{array}%
\right)$.
Notice that there are $(m-1)$ elements involved in the permutation.
This comes from the fact that the number of braid generators for $\B_n$
is given by $(n-1)$. This is also in consistent with
the cyclic symmetry for the color indices $c_i$.
As mentioned below (\ref{2-11}), four-dimensional diffeomorphism
is realized by permutation of the numbering indices in terms of $u_i$'s.
Invariance under diffeomorphism may be given by a {\it sum} over the permutation.
The notion of trace over braid generators, or the so-called braid trace,
therefore suggests diffeomorphism invariance for the holonomy operator.
This means that we can use a similar holonomy operator
for a gravitational theory as well.
The difference lies in the color degree of freedom, {\it i.e.}, the
so-called Chan-Paton factor.
In the spinor-momenta formalism, Lorentz invariance is manifest.
Thus, if we consider a certain representation of Poincar\'e algebra in this formalism,
the representation is essentially given by translational generators.
This suggests that a gravitational Chan-Paton factor
is described by a combination of the translational generators.
A gravitational analog of the holonomy operator
should contain a braid trace furnished with such a Chan-Paton factor,
which naturally leads to invariance under diffeomorphism.
We shall consider details of these points in the accompanying paper \cite{Abe:hol02}.

The braid generators satisfy the relation (\ref{3-2}). This is
equivalent to the Yang-Baxter equation if we neglect the notion of
spectral parameters.\footnote{
As we discuss in \cite{Abe:hol02}, an irreducible algebra of the
braid generators is given by the so-called Iwahori-Hecke algebra.
Thus we need to use this algebra
in order to relate the holonomy operator with physical quantities.
The Iwahori-Hecke algebra satisfies the braid relation (\ref{3-2})
on top of a certain condition on the generators.
Therefore, this algebra also contains the information of
Yang-Baxter equation and the following arguments do apply to
the irreducible case as well.
}
In considering the path ordering of the number indices,
we can in fact introduce a parameter $\tau$ ($0 \le \tau \le 1$) along the path.
Namely, we can parametrize the path $\ga$ on the physical configuration $\C$ as $\ga (\tau)$.
We then introduce a parameter $\tau_i$ corresponding to the index $i$ $(i = 1,2, \cdots, m$)
and require a condition $0 \le \tau_1 \le \tau_2 \le \cdots \le \tau_m \le 1$ according to
our choice of path ordering.
In this context, we can interpret a set of $\tau_i$'s as spectral parameters.
Thus, the introduction of ``path-ordered'' braid trace in (\ref{3-20}) means
the existence of Yangian symmetry in the holonomy operator.
(For recent investigation on Yangian symmetry
in scattering amplitudes of gluons, see, {\it e.g.},
\cite{Drummond:2009fd,Bargheer:2009qu}.)

From the expression (\ref{3-25}), we find that,
with a suitable normalization for
$\oint_\ga \om_{12} \wedge \cdots \wedge \om_{n1}$, the holonomy operator
can be used as a generating function for all physical states of gluons
on the Hilbert space $V^{\otimes n}$.
In order to extract physical quantities out of $\Theta_{R, \ga}$,
we have to express the operator in terms of spinor momenta $u_{i}^{A}$,
rather than $z_i$.
As mentioned before, below (\ref{2-8}), the differential
one-form $\om_{ij}$ can be expressed in terms of either $z_i$ or $u_{i}^{A}$.
This is because we can choose either the local coordinates
or the homogeneous coordinates as
covariant bases of a differential manifold on $\cp^1$.
We can therefore express the above results in terms of $u^A$'s as well.
We recapitulate these results below as a summary of this section.
\beqar
\Theta_{R, \ga} (u) &=& \Tr_{R, \ga} \, \Path \exp \left[
\sum_{m \ge 2} \oint_{\ga} \underbrace{A \wedge A \wedge \cdots \wedge A}_{m}
\right]
\label{3-28} \\
A &=&  \frac{1}{\kappa} \sum_{1 \le i < j \le n} A_{ij} \, \om_{ij}
\label{3-29} \\
\om_{ij} & = & d \log(u_i u_j) = \frac{d(u_i u_j)}{(u_i u_j)}
\label{3-30} \\
A_{ij} &=& a_{i}^{(+)} \otimes a_{j}^{(0)} + a_{i}^{(-)} \otimes a_{j}^{(0)}
\label{3-31}
\eeqar
For practical calculations, we also need the commutation relations (\ref{3-1}),
(\ref{3-14}) and the information about the color degree of freedom (\ref{3-21}).

\section{Supersymmetry and MHV amplitudes}

In the previous section, we focus on an algebraic structure of
the gauge field operators in twistor space.
The bialgebraic structure arises from
an analysis on the numbering indices of the gauge fields.
We now consider the other property of the gauge fields, {\it i.e.} the helicity of the particle.
Notice that, as discussed in (\ref{2-18}), the helicity of the particle is
determined by the degree of homogeneity in $u^A$'s.
In order to incorporate this information, we find that the most natural way is to
introduce $\N=4$ supersymmetry, as initially observed by Nair.\footnote{
Investigation of non-supersymmetric description to the spinor-momenta formalism
has been made considerably in preparing this paper
but every attempt turns out to be unsuccessful.
This is probably correlated with our specific choice of the ordering in (\ref{3-23}),
apart from the issue of conformal invariance.
}
So the twistor space $\cp^3$ now becomes the supertwistor space $\cp^{3|4}$.

\noindent
\underline{Supertwistor space}

Supertwistor space
is defined by the homogeneous coordinates of $\cp^{3|4}$,
$Z_{\hat{I}} = (u^A, v_\Ad, \xi^\al)$, where we introduce
Grassmann variables $\xi^\al = \th^{\al}_{A} u^A$ $(\al = 1,2,3,4)$
in addition to the twistor variables $Z_I =( u^A, v_\Ad)$ in (\ref{2-9}).
$I$ and $\widehat{I}$ are composite indices that can be
labeled as $I = 1,2,3,4$ and $\hat{I} = 1,2, \cdots , 8$, respectively.
Linear transformations of $Z_{\hat{I}}$ are accordingly represented by
the superconformal group $SU(2,2|4)$.
Coordinates of a compact four-dimensional spacetime $x_{\Ad A}$ and
their chiral (or holomorphic) superpartners $\th^{\al}_{A}$ arise
from supertwistor space with an imposition of the following relations.
\beq
v_\Ad = x_{A  \Ad} u^A \, , ~~~
\xi^\al = \th_{A}^{\al} u^A
\label{4-1}
\eeq
These are a supersymmetric analog of the twistor-space condition (\ref{2-10}).
As in the case of a superspace formalism, the coordinates $x_{\Ad A}$ can be
extended to $x_{\Ad A} \rightarrow x_{\Ad A} + 2 \bth_{\al \Ad} \th^{\al}_{A}$.
So a supersymmetric extension of the product $x_{\Ad A} p^{A \Ad}$ is expressed as
\beqar
x_{\Ad A} p^{A \Ad} & \rightarrow &
\left. x_{\Ad A} u^A \bu^\Ad +  2 \bth_{\al \Ad}  \th_{A}^{\al} u^A  \bu^\Ad
\right|_{v_\Ad = x_{\Ad A} u^A , \, \xi^\al = \th^{\al}_{A} u^A } \nonumber \\
&=& v_\Ad \bu^\Ad + 2 \bar{\et}_\al \xi^\al
\label{4-2}
\eeqar
where we impose the condition (\ref{4-1}) to the product. We also define antiholomorphic
Grassmann variables $\bar{\et}_\al$ ($\al = 1, 2, 3, 4$) as
\beq
\bar{\et}_\al = \bu_{\Ad}  \bth_{\al}^{\Ad}
\label{4-3}
\eeq

\noindent
\underline{Physical operators in $\Theta_{R, \ga} (u)$}

In this section, we consider a supersymmetric extension of the holonomy operator
$\Theta_{R, \ga} (u)$ in (\ref{3-28}) and relate the resultant quantity with
a generating functional for the maximally helicity violating (MHV) amplitudes of gluons.
Physical information of $\Theta_{R, \ga} (u)$ is encoded in
the operators $a_{i}^{(\pm)}$ of (\ref{3-31}); as discussed in the previous section,
we have teated $a_{j}^{(0)}$ of (\ref{3-31}) as
unphysical (or a kind of projective) operators.
Thus we should interpret $a_{i}^{(\pm)}$ as physical operators
and supersymmetrization should be made in them.
(Remember that the operators $a_{i}^{(\pm)}$ correspond to creation operators
of the $i$-th gluon with helicity $\pm$, which act on a
Fock space $V_i$ ($i = 1,2, \cdots , n)$. In terms of the $V_i$'s,
the physical Hilbert space of the spinor-momenta formalism is given by
$V^{\otimes n} = V_1 \otimes V_2 \otimes \cdots \otimes V_n$.)
In the spinor-momenta formalism,
there are two mutually related interpretations for $a_{i}^{(\pm)}$.
These can be stated as follows.
\begin{enumerate}
  \item The first approach is to interpret $a_{i}^{(\pm)}$ as operators in
a $\bu_i$-space representation, where a classical phase space is given by $(v_i, \bu_i)$.
Physical quantities in terms of the twistor variables $(u_i, v_i)$ are expressed
by a $v_i$-space representation,
or a Fourier transform (\ref{2-15}), of $a_{i}^{(\pm)}$.
  \item The second approach is to interpret $a_{i}^{(\pm)}$ as operators
in an ordinary $p_i$-space representation from the beginning,
where a classical phase space is given by $(x, p_i)$.
An $x$-space representation of $a_{i}^{(\pm)}$ is expressed by use of
the Nair measure (\ref{2-19}).
\end{enumerate}
Accordingly, there are two approaches towards a supersymmetric extension of
$\Theta_{R, \ga} (u)$.
In what follows, we discuss supersymmetrization of these approaches,
clarifying how $\N=4$ extended supersymmetry
realizes the helicity information of gluons.

\noindent
\underline{The first approach and Witten's interpretation for the MHV amplitudes}

In the first approach, we start with an interpretation
of the physical operators $a_{i}^{(\pm)}$ as operators labeled by
$\bu_i$. A $v_i$-space representation of $a_{i}^{(\pm)}$ can be
defined as
\beq
a_{i}^{(\pm)}(v_i) = \frac{1}{4}
\int \frac{d^2 \bu_i}{(2 \pi )^2} ~ a_{i}^{(\pm)} ~
e^{\frac{i}{2} \, v_{i \Ad} \bu_{i}^{\Ad}}
\label{4-4}
\eeq
This can be seen as a mode expansion $a_{i}^{(\pm)}(v_i)$
in terms of $\bu_i$-space wavefunctions.
Classically, we can regard $a_{i}^{(\pm)}$ as functions of $\bu_i$.
So the expression (\ref{4-4}) corresponds to a Fourier transform (\ref{2-15})
of such functions in twistor space.
Substituting $a_{i}^{(\pm)}(v_i)$ into $a_{i}^{(\pm)}$ of (\ref{3-31}),
and hence into (\ref{3-28}), we find that the holonomy operator
$\Theta_{R,\ga}(u)$ can be expressed as $\Theta_{R, \ga} (u; v)$.
With an imposition of the twistor condition $v_\Ad = x_{\Ad A} u^A$,
this is then expressed as $\Theta_{R, \ga} (u; x)$.
A supersymmetric extension of these procedures is carried out
by considering a supersymmetric version of (\ref{4-4}).

Before the supersymmetrization, let us first consider
the degree of homogeneity in $u_i$'s for
the holonomy operator $\Theta_{R, \ga}(u)$.
Notice that we take the logarithmic one-form $\om_{ij} = d \log (u_i u_j)$
as a covariant basis, which is what we previously call a propagator unit on $\cp^1$.
We then find that $\Theta_{R,\ga}(u)$ has zero degree of homogeneity in each of $u_i$'s.
Thus, from (\ref{2-18}), we can interpret that $\Theta_{R,\ga}(u)$
describes gluons of positive helicity.

As mentioned in the beginning of this section,
gluons of negative helicity can be incorporated by
an introduction of $\N = 4$ supersymmetry.
This can be seen as follows.
A phase space of interest now becomes $(v_i, \xi_i ; \bu_i, \bar{\et}_i)$.
Thus the operator $a_{i}^{(h_i)}$ in the $\bu_i$-space should be
expressed as $a_{i}^{(h_i)} (\bar{\et}_i)$, where $h_i$ denotes the helicity of
the $i$-th particle.
An operator labeled by $\xi_i$ can be defined as
\beq
a_{i}^{(h_i)} (\xi_i) = \int d^4 \bar{\et}_i ~ a_{i}^{(h_i)} (\bar{\et}_i ) ~
e^{i \, \bar{\et}_{i \al} \xi_{i}^{\al}}
\label{4-5}
\eeq
In the above expression, the degree of homogeneity in $u$'s arises only from the last factor
$\exp( i \, \bar{\et}_{i \al} \xi_{i}^{\al} ) = \exp( i \bar{\et}_i \cdot \xi_i )$
if we use the condition $\xi_{i}^{\al} = \th_{A}^{\al} u_{i}^{A}$.
Since we are considering $\N =4$ supersymmetry with $\al = 1,2,3,4=\N$,
this can be expanded as
\beq
\exp( i \bar{\et}_i \cdot \xi_i) = 1 + i \bar{\et}_i \cdot \xi_i +
\frac{1}{2!} (i \bar{\et}_i \cdot \xi_i)^2
+ \frac{1}{3!} (i \bar{\et}_i \cdot \xi_i )^3 + \frac{1}{4!} (i \bar{\et}_i \cdot \xi_i)^4
\label{4-6}
\eeq
which gives an $\N = 4 $ supermultiplet.
The first and the last terms correspond to gluons of positive and negative helicities,
respectively. The remaining terms can be interpreted as the superpartners
of gluons, corresponding to gluinos with spin-$\hf$ and to a scalar particle.
The helicity for the $i$-th particle or supermultiplet described by
$a_{i}^{(h_i )} (\xi_i)$ is then expressed as
\beq
h_i = 1 - \hf \sum_\al \xi_{i}^{\al} \frac{\d}{\d \xi_{i}^{\al}}
\label{4-7}
\eeq
Thus we find that the number of $\xi_i$'s in (\ref{4-5}) determines the helicity.
Notice that the use of Grassmann variables is
crucial for the truncation of the expansion (\ref{4-6}).
Further, the Grassmann integral in (\ref{4-5}) vanishes except in the case of
\beq
\int d^4 \bar{\et}_i ~ \bar{\et}_{i1} \bar{\et}_{i2}
\bar{\et}_{i3}\bar{\et}_{i4} = 1
\label{4-8}
\eeq
Utilizing this relation, we can suitably define
$a_{i}^{(h_i)} (\bar{\et}_i )$ in (\ref{4-5})
such that $a_{i}^{(h_i)} (\xi_i)$ are expressed as
\beqar
\nonumber
a_{i}^{(+)} (\xi_i) &=& a_{i}^{(+)} \\ \nonumber
a_{i}^{\left( + \hf \right)} (\xi_i) &=& \xi_{i}^{\al}
\, a_{i \al}^{ \left( + \hf \right)} \\
a_{i}^{(0)} (\xi_i) &=& \hf \xi_{i}^{\al} \xi_{i}^{\bt} \, a_{i \al \bt}^{(0)} \label{4-9}
\\ \nonumber
a_{i}^{\left(- \hf \right)} (\xi_i) &=& \frac{1}{3!} \xi_{i}^{\al}\xi_{i}^{\bt}\xi_{i}^{\ga}
\ep_{\al \bt \ga \del} \, {a_{i}^{ \del}}^{ \left( - \hf \right)}
\\ \nonumber
a_{i}^{(-)} (\xi_i) &=& \xi_{i}^{1} \xi_{i}^{2} \xi_{i}^{3} \xi_{i}^{4} \, a_{i}^{(-)}
\eeqar
where the number of $\xi_i$'s in each of $a_{i}^{(h_i)} (\xi_i)$
respects the helicity relation (\ref{4-7}).
Thus we find that an introduction of $\N = 4$ supersymmetry to
the physical operators $a_{i}^{(h_i)}$ naturally leads to
the description of gluons with both positive and negative helicities.

Operators labeled by the ``coordinate'' variables $(v_i , \xi_i)$ are
easily written as
\beq
a_{i}^{(h_i)} (v_i , \xi_i)
=  \frac{1}{4} \int \frac{d^2 \bu_i}{(2 \pi)^2} \, a_{i}^{(h_i)} (\xi_i)
\, e^{\frac{i}{2} \, v_{i \Ad} \bu_{i}^{\Ad}}
\label{4-10}
\eeq
which gives a supersymmetric extension of (\ref{4-4}).
Notice that, at a function level, (\ref{4-10}) corresponds to
a Fourier transform of $a_{i}^{( h_i )}$ in supertwistor space.

{\it
A supersymmetric holonomy operator $\Theta_{R, \ga} (u; v, \xi)$ is defined by
substitution of $a_{i}^{(h_i)} (v_i , \xi_i)$ into $a_{i}^{(h_i)}$ $(h_i = \pm )$
in $(\ref{3-31})$, and hence into the expression
of $\Theta_{R, \ga} (u)$ in $(\ref{3-28})$.
}

The MHV tree amplitudes for gluons are the scattering amplitudes of
$(n-2)$ gluons of positive helicity and $2$ gluons of negative
helicity.
In a momentum-space representation, the amplitudes are expressed
in terms of $(u_i , \bu_i)$ as
\beqar
\nonumber
\A_{MHV}^{(1_+ 2_+ \cdots r_{-} \cdots s_{-} \cdots n_+ )} (u, \bu)
& \equiv &
\A_{MHV}^{(r_{-} s_{-})} (u, \bu) \\
& = & i g^{n-2}
\, (2 \pi)^4 \del^{(4)} \left( \sum_{i=1}^{n} p_i \right) \,
\widehat{A}_{MHV}^{(r_{-} s_{-})} (u)
\label{4-11}
\eeqar
\beq
\widehat{A}_{MHV}^{(r_{-} s_{-})} (u) =
\sum_{\si \in \S_{n-1}}
\Tr (t^{c_1} t^{c_{\si_2}} t^{c_{\si_3}} \cdots t^{c_{\si_n}}) \,
\frac{ (u_r u_s )^4}{ (u_1 u_{\si_2})(u_{\si_2} u_{\si_3})
\cdots (u_{\si_n} u_1)}
\label{4-12}
\eeq
where $g$ is the coupling constant and the labels
$r$ and $s$ denote the numbering indices
of the negative-helicity gluons.
$t^{c_i}$'s are the generators of the gauge group in the
$R$-representation. The sum of $\S_{n-1}$ is taken over the
permutations of the elements $\{2,3, \cdots, n \}$.

We can construct a generating functional for the MHV gluon amplitudes
in terms of the supersymmetric holonomy  operator $\Theta_{R, \ga} (u; v, \xi)$ as
\beq
\F_{MHV} \left[ a^{(h)c} \right] = \exp \left[ \frac{i}{g^2}
\int  d^8 \th ~ (2 \pi)^4 \del^{(4)} \left( \sum_{i=1}^{n} p_i \right) ~
\Theta_{R, \ga} (u; v, \xi)
\right]_{\xi^\al = \th^{\al}_{A} u^A }
\label{4-13}
\eeq
where $a^{(h)c}$ refers to a generic expression for $a_{i}^{(h_i)c_i}$
$(i = 1,2, \cdots )$, with $a_{i}^{(h_i)}$ being
$a_{i}^{(h_i)} =  t^{c_i} \, a_{i}^{(h_i)c_i}$ as in (\ref{3-21}).
From equation (\ref{3-29}), we find that the coupling constant $g$
can be parametrized as
\beq
g = \frac{1}{\kappa}
\label{4-14}
\eeq
where $\kappa$ is a constant which appears in the definition of
the KZ equation (\ref{3-6}). We shall discuss the significance of
this relation later.

In terms of the functional $\F_{MHV} \left[ a^{(h)c} \right]$,
the MHV amplitudes are generated as
\beqar
\nonumber
&& \!\!\!\!\!\!\!
\left. \frac{\del}{\del a_{1}^{(+) c_1}} \otimes
\cdots \otimes \frac{\del}{\del a_{r}^{(-) c_r}} \otimes
\cdots \otimes \frac{\del}{\del a_{s}^{(-) c_s}} \otimes
\cdots \otimes \frac{\del}{\del a_{n}^{(+) c_n}}
~ \F_{MHV} \left[  a^{(h)c} \right] \right|_{a^{(h)c}=0} \\
\nonumber
&=&
i g^{n-2} (2 \pi)^4 \del^{(4)} \left( \sum_{i=1}^{n} p_i \right)
\,
\left[
\prod_{i=1}^{n} \frac{1}{4} \int \frac{d^2 \bu_i}{(2 \pi)^2}
\, e^{\frac{i}{2} \,  v_{i \Ad} \bu_{i}^{A} }
\right]
\\
\nonumber
&& ~~~~~~~~~~~~~~~ \times \, \Tr (t^{c_1}t^{c_{2}} \cdots t^{c_{n}})
\frac{(u_r u_s)^4}{(u_1 u_2)(u_2 u_3) \cdots (u_n u_1)}
~ + ~ \P(2,3, \cdots, n) \\
\nonumber
&=&
\left[
\prod_{i=1}^{n} \frac{1}{4} \int \frac{d^2 \bu_i}{(2 \pi)^2}
~ e^{\frac{i}{2} \, v_{i \Ad} \bu_{i}^{A} }
\right]
~ \A_{MHV}^{(r_{-}, s_{-})} (u, \bu)
\\
&=&
\A_{MHV}^{(r_{-}, s_{-})} (u, v)
\label{4-15}
\eeqar
where $\P (2,3, \cdots , n)$ indicates
the terms obtained by permutations of the indices
$\{ 2,3,\cdots , n \}$.
The condition $a^{(h)} = 0$ means that the remaining operators (or
source functions in a context of functional derivatives) should be evaluated as
zero in the end of the calculation.
In the above expression, we choose the normalization of
a holomorphic factor as
\beq
\oint_\ga d(u_1 u_2) \wedge d(u_2 u_3) \wedge \cdots \wedge d (u_n u_1) = 2^{n+1}
\label{4-16}
\eeq
From (\ref{3-25}), we find that this is a proper choice for the
derivation of scattering amplitudes from $\Theta_{R , \ga} (u)$.
We omit an irrelevant sign factor $(-1)^{h_1 + h_2 + \cdots + h_n}$ in
(\ref{4-15}) since physical quantities are given by the squares of the amplitudes.
Notice that the Grassmann integral over $\th$'s picks up only the MHV amplitudes
since the integral vanishes except when we have the following factor.
\beq
\left. \int d^8 \th  \, \xi_{r}^{1}\xi_{r}^{2}\xi_{r}^{3}\xi_{r}^{4}
\, \xi_{s}^{1}\xi_{s}^{2}\xi_{s}^{3}\xi_{s}^{4}
\right|_{\xi_{i}^{\al} = \th_{A}^{\al} u_{i}^{A} }
= (u_r u_s )^4
\label{4-17}
\eeq
We therefore find that the supersymmetric holonomy operator $\Theta_{R, \ga}(u; v, \xi)$
naturally describes a generating functional for the gluon MHV amplitudes.
Appearance of integrals over $\bu_i$'s in (\ref{4-15}) indicates that we have
interpreted the operators $a_{i}^{(h_i)}$ as those in the $\bu_i$-spaces.
This arises from our choice of phase spaces being $(v_i , \bu_i)$ as
mentioned before.
The MHV tree amplitudes (\ref{4-11}) as functions of $(u, \bu)$
are obtained from (\ref{4-15}) by inverse Fourier transforms in twistor space.
\beq
\A_{MHV}^{(r_- s_-)} (u, \bu) \, = \,
\left[
\prod_{i=1}^{n} \int d^2 v_i
~ e^{-\frac{i}{2} \, v_{i \Ad} \bu_{i}^{\Ad} }
\right]
~\A_{MHV}^{(r_{-}, s_{-})} (u, v)
\label{4-18}
\eeq

It is essentially this line of expressions
that has been developed by Witten for the discussion of
the MHV amplitudes in \cite{Witten:2003nn},
while in Witten's original approach the supertwistor condition
$\xi_{i}^{\al} = \th_{A}^{\al} u_{i}^{A}$
is not imposed by hand; it is built in the definition of the MHV amplitudes.
Following Witten's prescription, we now introduce the following
fermionic delta-function.
\beq
\del^{(8)} \left( \sum_{i=1}^{n} p_{i \, \Ad}^{A} \bth_{\al}^{\Ad} \right)
= \del^{(8)} \left( \sum_{i=1}^{n} u_{i}^{A} \bar{\et}_{i \al}  \right)
= \int d^8 \th ~ e^{- i \, \th_{A}^{\al}
\sum_i u_{i}^{A} \bar{\et}_{i \al}}
\label{4-19}
\eeq
In terms of this delta-function, we can define a generating functional
for the MHV amplitudes such that the condition
$\xi_{i}^{\al} = \th_{A}^{\al} u_{i}^{A}$
is automatically embedded in the resultant amplitudes.
In analogy with (\ref{4-13}), such a functional can be expressed as
\beq
\F_{MHV} \left[ a^{(h)c} (\bar{\et}) \right]
= \exp \left[
\frac{i}{g^2} (2 \pi)^4 \del^{(4)} \left( \sum_i p_i \right)
\, \del^{(8)} \left( \sum_i u_{i}^{A} \bar{\et}_{i \al} \right)
~ \Theta_{R, \ga} (u; v, \xi)
\right]
\label{4-20}
\eeq
where $a^{(h)c} (\bar{\et})$ refers to a generic expression for
$a_{i}^{(h_i)c_i} (\bar{\et}_i)$, with $a_{i}^{(h_i)} (\bar{\et}_i)$
being written as
$a_{i}^{(h_i)} (\bar{\et}_i) = t^{c_i} a_{i}^{(h_i)c_i} (\bar{\et}_i)$s;
notice that $a_{i}^{(h_i)} (\bar{\et}_i)$'s are defined in (\ref{4-5}).

Utilizing $\F_{MHV} \left[ a^{(h)c} (\bar{\et}) \right]$ and
the relation (\ref{4-19}), we can generate the MHV amplitudes as
\beqar
\nonumber
&& \!\!\!\!\!\!\!
\left. \frac{\del}{\del a_{1}^{(+) c_1}( \bar{\et}_1 )} \otimes
\cdots \otimes \frac{\del}{\del a_{r}^{(-) c_r}( \bar{\et}_r )} \otimes
\cdots
~ \F_{MHV} \left[ a^{(h)c} ( \bar{\et} ) \right]
 \right|_{a^{(h)c} ( \bar{\et} ) =0} \\
\nonumber
&=&
i g^{n-2} \left[ \prod_{i=1}^{n}  \frac{1}{4} \int
\frac{d^2 \bu_i}{(2 \pi)^2} d^4 \bar{\et}_i \right]
\int d^4 x d^8 \th ~ \exp \left(
- \frac{i}{2}  x_{\Ad A} \sum_i u_{i}^{A} \bu_{i}^{\Ad}
-i  \th_{A}^{\al} \sum_i u_{i}^{A} \bar{\et}_{i \al}
\right) \\
\nonumber
&& ~~~~~~~~~~~~~~~~~~~~~~~~~~~
\times \, \exp \left(
\frac{i}{2} \sum_{i} v_{i \Ad} \bu_{i}^{\Ad}
+ i \sum_{i} \bar{\et}_{i \al} \xi_{i}^{\al}
\right)
~ \widehat{A}_{MHV}^{(r_- s_-)} (u) \\
\nonumber
&=&
\int d^4 x d^8 \th \,
\prod_{i=1}^{n} \del^{(2)} ( v_{i \Ad} - x_{\Ad A} u_{i}^{A} ) \,
\del^{(4)} (\xi_{i}^{\al} - \th_{A}^{\al} u_{i}^{A} ) ~
\widehat{A}_{MHV}^{(r_- s_-)} (u) \\
& \equiv &
\A_{MHV}^{(r_{-}, s_{-})} (u, v, \xi)
\label{4-21}
\eeqar
where we use the relation
\beq
(2 \pi)^4 \del^{(4)} \left( \sum_i p_i \right) =
\int d^4 x ~ e^{- i \, x_\mu \sum_i p_{i}^{\mu}}
= \int d^4 x ~ e^{- \frac{i}{2} \, x_{\Ad A} \sum_i u_{i}^{A} \bu_{i}^{\Ad}}
\label{4-22}
\eeq
The resultant amplitudes relate to the MHV amplitudes
$\A_{MHV}^{(r_- s_-)} (u, \bu)$ in (\ref{4-11}) by
\beqar
\nonumber
&& \!\!\!\!\!\!
\left[
\prod_{i=1}^{n}
\int d^2 v_i d^4 \xi_i
\, e^{-\frac{i}{2} v_{i \Ad} \bu_{i}^{\Ad}} \,
e^{- i \bar{\et}_{i \al} \xi_{i}^{\al} }
\right] ~ \A_{MHV}^{(r_- s_-)} (u,v, \xi)
 \\
&=& \A_{MHV}^{(r_- s_-)} (u, \bu, \bar{\et})
\, = \,
\del^{(8)} \left( \sum_i u_{i}^{A} \bar{\et}_{i \al} \right)
\, \A_{MHV}^{(r_- s_-)} (u, \bu)
\label{4-23}
\eeqar
where the first line corresponds to
inverse Fourier transforms in supertwistor space.

Equation (\ref{4-21}) shows that
the supertwistor condition (\ref{4-1}) is automatically embedded in
the resultant amplitudes.
This expression arises from the
attachment of the supersymmetric delta-function to
a function of supertwistor variables, or
what we can define as $\widehat{A}_{MHV}^{(r_- s_-)} (u, v, \xi)$,
{\it without} an imposition of the supertwistor condition.
Thus, in Witten's approach, we can regard the supertwistor
condition as an emergent relation.
For a given set of $(x, \th)$, this condition determines a curve in
supertwistor space. This curve can be characterized by the
spinor momenta $u_{i}^{A}$ since we may solve the condition in terms of
$u_{i}^{A}$.
The four-dimensional chiral super-coordinates $(x, \th)$ are then interpreted
as moduli of this curve.

\noindent
\underline{The second approach and Nair's observation for the MHV amplitudes}

The second approach is more straightforward in terms of extracting four-dimensional
information since we start with an interpretation of the operators $a_{i}^{(h_i)}$
as those in an ordinary four-dimensional $p_i$-space representation.
An $x$-space representation of the operators is given by
\beq
a_{i}^{(h_i)} (x) =
\int d \mu (p_i)
\,  a_{i}^{(h_i)} \,  e^{i x_\mu p_{i}^{\mu} }
\label{4-24}
\eeq
where $d \mu (p_i)$ denotes the Nair measure (\ref{2-19}).
By use of the extension
$x_{\Ad A} \rightarrow x_{\Ad A} + 2 \bth_{\al \Ad} \th^{\al}_{A}$,
a supersymmetric extension of these operators can be obtained as
\beq
a_{i}^{(h_i)} (x, \th)
\, = \,
\left. \int d\mu (p_i) ~ a_{i}^{(h_i)} (\xi_i) ~  e^{ i x_\mu p_{i}^{\mu} }
\right|_{\xi_{i}^{\al} = \th_{A}^{\al} u_{i}^{A} }
\label{4-25}
\eeq
where $a_{i}^{(h_i)} (\xi_i)$'s are defined in (\ref{4-9}). Notice that
we impose the supertwistor condition $\xi_{i}^{\al} = \th_{A}^{\al} u_{i}^{A}$ as in
(\ref{4-13}).

{\it
A supersymmetric holonomy operator $\Theta_{R, \ga} (u; x, \th)$ is defined by
substitution of $a_{i}^{(h_i)} (x , \th)$ into $a_{i}^{(h_i)}$ $(h_i = \pm )$
in $(\ref{3-31})$, and hence into the expression
of $\Theta_{R, \ga} (u)$ in $(\ref{3-28})$.
}

In terms of $\Theta_{R, \ga} (u; x, \th)$, we can construct an S-matrix functional
for the MHV gluon amplitudes as
\beq
\F_{MHV} \left[ a^{(h)c} \right]
 = \exp \left[ \frac{i}{g^2} \int d^4 x d^8 \th ~ \Theta_{R, \ga} (u; x ,\th) \right]
\label{4-26}
\eeq
where $a^{(h)c}$ now denotes a generic expression for $a_{i}^{(h_i)c_i}$, with
$a_{i}^{(h_i)} = t^{c_i} a_{i}^{(h_i)c_i}$ understood as operators in
the $p_i$-space representation.
Since the operator is a function of $(u; x, \th)$,
a relevant form of the MHV amplitudes is given in an $x$-space
representation as
\beq
\A_{MHV}^{(1_+ 2_+ \cdots r_{-} \cdots s_{-} \cdots n_+ )} (x)
\equiv
\A_{MHV}^{(r_{-} s_{-})} (x)
 =
\prod_{i=1}^{n} \int d \mu (p_i) ~ \A_{MHV}^{(r_{-} s_{-})} (u, \bu)
\label{4-27}
\eeq
where $\A_{MHV}^{(r_{-} s_{-})} (u, \bu)$ is defined in (\ref{4-11}).
In terms of (\ref{4-26}), this MHV amplitudes can be expressed as
\beqar
\nonumber
&& \!\!\!\!\!\!\!
\left. \frac{\del}{\del a_{1}^{(+) c_1}} \otimes
\cdots \otimes \frac{\del}{\del a_{r}^{(-) c_r}} \otimes
\cdots \otimes \frac{\del}{\del a_{s}^{(-) c_s}} \otimes
\cdots \otimes \frac{\del}{\del a_{n}^{(+) c_n}}
~ \F_{MHV} \left[  a^{(h)c} \right] \right|_{a^{(h)c}=0} \\
&=&
\A_{MHV}^{(r_{-} s_{-})} (x)
\label{4-28}
\eeqar
where we should emphasize that $\F_{MHV} \left[ a^{(h)c} \right]$ is defined
in (\ref{4-26}) and that the operators $a_{i}^{(h_i)c_i}$, which are treated as
source functions here, are labeled by the four-dimensional momenta $p_i$.
The above expression gives an explicit realization of
Nair's original observation to the MHV amplitudes \cite{Nair:1988bq}.
This form is also obtained in \cite{Abe:2004ep}.
Notice that, in terms of the holomorphic amplitudes
$\widehat{A}_{MHV}^{(r_{-} s_{-})} (u)$ in (\ref{4-12}),
the expression can be simplified as
\beqar
\nonumber
&& \!\!\!\!\!\!\!
\frac{\del}{\del a_{1}^{(+) c_1} (x_{1}) } \otimes
\cdots \otimes \frac{\del}{\del a_{r}^{(-) c_r} (x_{r})} \otimes \cdots
\\
\nonumber
&& ~~~~~~~~~~~~~~~~
\left. \cdots \otimes \frac{\del}{\del a_{s}^{(-) c_s} (x_{s})} \otimes
\cdots \otimes \frac{\del}{\del a_{n}^{(+) c_n} (x_{n})}
~ \F_{MHV} \left[ a^{(h)c} \right]
 \right|_{a^{(h)c} ( x ) =0}
\\
& = &
i g^{n-2} \, \widehat{A}_{MHV}^{(r_{-} s_{-})} (u)
\label{4-29}
\eeqar
where $a^{(h)c} ( x )$'s play the same role as $a^{(h)c}$'s in (\ref{4-28}) except that
they are now in an $x$-space representation as in (\ref{4-24}).

In Nair's observation, the MHV gluon amplitudes arise from a current correlator
of a WZW model. In the present formalism, the current structure arises from
our choice of logarithmic one-forms $\om_{ij}$ in (\ref{2-8})
and the subsequent construction of $n$-forms in terms of the
comprehensive gauge one-form $A$ in (\ref{3-16}).
Current correlators of a WZW model satisfy the KZ equation in general.
In the present formalism, the current correlators can be defined as
functions on the configuration space $\C = {\bf C}^n / \S_n$ and the
KZ equation corresponds to the requirement of the functions being
covariantly constant, where a covariant derivative is defined in
terms of $A$, as discussed in the previous section.
By use of the operator product expansion (OPE) of a WZW current,
it is argued in \cite{Nair:1988bq} that
the level number of the WZW model should be fixed to $k = 1$
in order to interpret the correlators as the MHV amplitudes.
The choice of $k = 1$ is also appropriate because
it leads to an integrable representation of spin-1 particles.
In the present holonomy formalism, this fact should be reflected in
the choice of the KZ parameter $\kappa$,
or the Yang-Mills coupling constant $g$ if we use the relation (\ref{4-24}).
For an $SU(N)$ gauge group, it is well known that the KZ parameter can be
given by
\beq
\kappa = k + N
\label{4-30}
\eeq
(For this relation, see, {\it e.g.},  \cite{CFTbook}.)
In the present case, a relevant group seems to be $SL(2, {\bf C}) = SU(2)^{{\bf C}}$,
where $SU(2)^{{\bf C}}$ denotes the complexification of $SU(2)$.
In the spinor-momenta formalism, holomorphicity plays a crucial role.
In fact, we begin with the complex Riemann surface $\cp^1$
to define holomorphic and antiholomorphic spinor momenta.
As shown in (\ref{4-11}), physical quantities may be described
by holomorphic functions, such as $\widehat{A}_{MHV}^{(r_{-} s_{-})} (u)$,
in terms of these spinor momenta.
If we make the energy-momentum conservation explicit as in (\ref{4-11}),
then we need to deal with full complex functions such as
$\A_{MHV}^{(r_{-} s_{-})} (u, \bu)$.
Thus, in order to define gauge transformations on
such complex functions, it is natural to extend
the gauge group $SU(N)$ to the complexified gauge group
$SL(N, {\bf C})$.
Since any $SL(N, {\bf C})$ group contains an $SL(2, {\bf C})$ as a subgroup,
we can consider that a physically relevant KZ parameter
is given by (\ref{4-30}).\footnote{
At the present, it is under investigation to show this relation
in a more rigorous manner.}
Therefore, we can parametrize the coupling constant as
\beq
g = \frac{1}{1 + N}
\label{4-31}
\eeq
Since four-dimensional coupling constants are dimensionless, this
is an appropriate form. The significance of this result is that it
indicates a nonperturbative nature of the present formalism;
notice that the $N$ dependence of $g$ is conventionally
considered as $g \sim 1 / \sqrt{N}$ either in
a large $N$ analysis of Yang-Mills theory or in perturbative string theory.


In this section, we have discussed a supersymmetric
extension of the holonomy operator $\Theta_{R, \ga} (u)$, which
naturally generates the MHV amplitudes of gluons.
There are two approaches towards the supersymmetrization of
$\Theta_{R, \ga} (u)$, corresponding to
different approaches to the MHV amplitudes taken by Nair and Witten,
respectively \cite{Nair:1988bq,Witten:2003nn}.
As mentioned before, the difference lies in the choice of
a classical phase space in calculations of the MHV amplitudes.
In Witten's approach, the supertwistor condition (\ref{4-1}) emerges
automatically so that physical quantities
are not necessarily constrained by (\ref{4-1}).
This leads to the idea of algebraic curves in $\cp^{3 | 4}$
which we have briefly mentioned in the introduction.
Non-MHV amplitudes of gluons are then classified by these algebraic curves.
This approach is therefore mathematical and abstract
in a sense that a framework of extracting
four-dimensional spacetime is beyond Penrose's original idea of twistor space.
On the other hand,
Nair's approach is conceptually more close to Penrose's idea.
In Nair's approach, the supertwistor condition
is imposed by hand on a function of the supertwistor variables.
Four-dimensional spacetime is then emerged \'a la Penrose.
In order to describe non-MHV amplitudes in this approach,
however, one needs to resort to the so-called Cachazo-Svrcek-Witten (CSW) rules
which we consider in the next section.

\section{CSW prescription and non-MHV amplitudes}

The generating functional $\F_{MHV}$ in (\ref{4-26})
literally generates only the MHV amplitudes. As seen above, this is
due to the $\N = 4$ Grassmann integral acting on the
supersymmetric holonomy operator $\Theta_{R, \ga} (u; x, \th)$.
In this section, we consider generalization of the above
formulation to non-MHV amplitudes at the tree level.

\noindent
\underline{CSW rules and a contraction operator}

One idea is to express the non-MHV amplitudes
in terms of a combination of the MHV amplitudes.
This is the idea developed by Cachazo, Svrcek and Witten in \cite{Cachazo:2004kj};
it has been shown that non-MHV amplitudes can be expressed by
a combination of MHV amplitudes (or vertices) connected with off-shell propagators.
A momentum transfer, say $q$, for each of the propagators is given by a sum of
the momenta involved in the corresponding MHV vertex.
The propagator is then realized simply by an inverse square of $q$
in a momentum-space representation of the non-MHV amplitudes.
This prescription for the non-MHV amplitudes is called the CSW rules.

In a language of functional derivative, this prescription suggests an
introduction of a contraction operator to $\F_{MHV}$.
In order to define such a contraction operator, we need to
introduce a propagator that connects two holonomy operators,
say $\Theta_{R, \ga} (u; x, \th)$ and $\Theta_{R, \ga} (u; y, \th)$.
The propagator connects the coordinates $x$ and $y$, each representing an MHV vertex.
The coordinates $(x,y)$ should be characterized by a set of
positive and negative helicities, not just by positive or negative helicities,
since the purpose of introducing the propagator is to increase the number of
negative-helicity states involved in scattering of interest
by expressing the scattering amplitudes in terms of a combination of MHV vertices.
In other words, we consider the propagator as an analog of a propagator
for complex scalar fields.

In what follows, we simply {\it restate} the CSW rules
by use of the expression (\ref{4-29}),
where the physical information is encoded in the operators $a_{i}^{(h_i)}(x)$
in an $x$-space representation.
The non-MHV amplitudes are then
described by insertions of a pair of positive and negative
helicity states, say $a_{k}^{(+)}(x)$ and $a_{l}^{(-)}(y)$,
into the original sequence of indices $(1, 2, \cdots, n)$ with
some concrete non-MHV helicity information. The number of insertions
represents how the helicity configuration of interest deviates from the
MHV configurations, which is equivalent to the number of negative-helicity states
minus two.

For simplicity of discussion, we first consider the next-to-MHV (NMHV) amplitudes
where there are three negative-helicity states labeled by, say, $r_{-}$, $s_{-}$ and $t_{-}$
($r < s < t$), with the rest of indices having positive helicities.
Assume that $r_{-}$ and $s_{-}$ are in between $i$ and $j$, {\it i.e.},
$1 \le i < r < s < j < t \le n$, then possible ways to split the NMHV amplitudes into two
MHV amplitudes can be written as
\beqar
\nonumber
&& 1_+ 2_+ \cdots \underbrace{i_+ \cdots r_- \cdots s_- \cdots j_+}_{\Rightarrow
~ q_{ij} = p_i + p_{i+1} \cdots + p_j}
 \cdots t_- \cdots n_+ \\
\label{5-1}
&& \Rightarrow ~
\underbrace{i_+ \cdots r_- \cdots s_- \cdots j_+ \, k_+}_{\widehat{A}_{MHV}^{(r_{-} s_{-})}(u)}
\underbrace{~~~~}_{1 / q_{ij}^{2}}
\underbrace{l_- \, (j+1)_{+} \cdots t_- \cdots (i-1)_{+}}_{\widehat{A}^{(l_{-} t_{-})}_{MHV}(u)}
\eeqar
where in the second line we schematically show the CSW rules, with
the factor of $1 / q_{ij}^{2}$ representing a propagator that connects the
two holomorphic MHV amplitudes $\widehat{A}_{MHV}^{(r_{-} s_{-})}(u)$
and $\widehat{A}_{MHV}^{(l_{-} t_{-})}(u)$.
Notice that an insertion of the indices $(k_+ , l_-)$ naturally leads to
separation of an NMHV amplitude into two MHV amplitudes, given
that the number of index arguments for each of the MHV amplitudes
is more than two.
Under this restriction, the CSW rules for the NMHV amplitudes
can be expressed as
\beq
\widehat{A}^{(r_- s_- t_-)}_{NMHV} (u) = \sum_{(i,j)}
\widehat{A}^{(i_+ \cdots r_- \cdots s_- \cdots j_+ k_+)}_{MHV} (u)
\, \frac{\del_{kl}}{q_{ij}^2} \,
\widehat{A}^{(l_- \, (j+1)_+ \cdots t_- \cdots (i-1)_+)}_{MHV} (u)
\label{5-2}
\eeq
where the sum is taken over all possible choices for $(i, j)$ that
satisfy the ordering $i < r < s < j < t$. The momentum transfer
$q_{ij}$ is given by
\beq
q_{ij} = p_i + p_{i+1 } + \cdots + p_{r} + \cdots + p_{s} + \cdots + p_{j}
\label{5-3}
\eeq
where $p$'s denote four-momenta of gluons as before.

A generating functional for the NMHV amplitudes (\ref{5-2}) is then
expressed as
\beqar
\F \left[ a^{(h)c} \right] &= &  \widehat{W} \, \F_{MHV} \left[ a^{(h)c} \right]
\label{5-4}\\
\widehat{W} &=& \exp \left[
\int d^4 x d^4 y ~ \frac{\del_{kl}}{q^2} ~
\frac{\del}{\del a_{k}^{(+)}(x)} \otimes
\frac{\del}{\del a_{l}^{(-)}(y)}
  \right]
\label{5-5} \\
\F_{MHV} \left[ a^{(h)c} \right]
& = & \exp \left[ \frac{i}{g^2} \int d^4 x d^8 \th
~ \Theta_{R, \ga} (u; x ,\th) \right]
\label{5-6}
\eeqar
where we rewrite $\F_{MHV}$ for convenience.
By use of $\F$ in (\ref{5-4}), we can generate the NMHV amplitudes as
\beqar
\nonumber
&& \!\!\!\!\!
\frac{\del}{\del a_{1}^{(+) c_1} (x_1) } \otimes
\cdots \otimes \frac{\del}{\del a_{r}^{(-) c_r} (x_r)} \otimes
\cdots \otimes \frac{\del}{\del a_{s}^{(-) c_s} (x_s)} \otimes
\cdots \\
\nonumber
&& ~~~~~~~~~~~~~~~
\left. \cdots \otimes \frac{\del}{\del a_{t}^{(-) c_t} (x_t)} \otimes
\cdots \otimes \frac{\del}{\del a_{n}^{(+) c_n} (x_n)}
~ \F \left[  a^{(h)c}  \right] \right|_{a^{(h)c} (x)=0} \\
&=&
i g^{n-2} \widehat{A}^{(r_- s_- t_-)}_{NMHV} (u)
\label{5-7}
\eeqar
Notice that the sum over $(i,j)$ in (\ref{5-2}) is automatically
realized by the functional derivatives acting on $\F$
and by the Grassmann integrals executed in $\F$.
This explains that the momentum transfer is denoted by
$q$ without the $(i, j)$ indices in (\ref{5-5}).
It is interesting that, in these expressions,
there is no appearance of $d \mu (q)$, or an integration measure
for the off-shell momentum.
Thus, we need no auxiliary fields to describe the
non-MHV amplitudes with the apparent off-shell propagators;
we simply need the physical operators
$a_{i}^{(h_i)} (x) = t^{c_i} a_{i}^{(h_i)c_i} (x)$
for gluons (which by definition include Nair measures)
in order to describe physical quantities generated by $\F$.
Another important point is that the
color structure of $\widehat{A}^{(r_- s_- t_-)}_{NMHV} (u)$ is
properly written as
$\Tr (t^{c_1} t^{c_{\si_2}} t^{c_{\si_3}} \cdots t^{c_{\si_n}} )$
with permutation over
$\si=\left(%
\begin{array}{c}
  2 \, 3 \cdots n \\
  \si_2 \si_3 \cdots \si_n \\
\end{array}%
\right)$
as shown in (\ref{4-12}).
This is guaranteed by Kronecker's delta $\del_{kl}$ and by the
fact that $t^{c_k}$ or $t^{c_l}$ can be considered as
a $U(1)$ direction which can be attached to the color degrees of
freedom. This $U(1)$ symmetry is related to the
phase invariance for spinor momenta in (\ref{2-3}).
This is also related to the
cyclic invariance for the trace of color structure, or the
so-called Chan-Paton factor of Yang-Mills theory.

\noindent
\underline{Non-MHV amplitudes}

We now consider the generalization to other non-MHV amplitudes which
are usually referred to as the N$^{m}$MHV amplitudes ($m = 1,2, \cdots , n-4$), with
$m+2$ being the number of negative-helicity gluons.
For the maximal case ($m = n-4$), the amplitudes become the so-called googly MHV amplitudes.
From our construction, we find that
they can be obtained by taking conjugates of the MHV amplitudes.
We notice that generalization of the expression (\ref{5-7}) to the N$^{m}$MHV amplitudes
is straightforward since
the Wick-like operator (\ref{5-5}) can be used for the
arbitrary number of contractions acting on $\F_{MHV}$.
Therefore, the functional $\F$ in (\ref{5-4}) can also be
used as an S-matrix functional for the non-MHV amplitudes in general.
An explicit expression can be written as
\beqar
&&
\nonumber
\!\!\!\!\!\!\!
\left. \frac{\del}{\del a_{1}^{(h_1) c_1} (x_1) } \otimes
\frac{\del}{\del a_{2}^{(h_2) c_2} (x_2)} \otimes
\cdots \otimes \frac{\del}{\del a_{n}^{(h_n) c_n} (x_n)}
~ \F \left[  a^{(h)c}  \right] \right|_{a^{(h)c} (x)=0} \\
&=&
i g^{n-2} \widehat{A}^{(1_{h_1} 2_{h_2} \cdots n_{h_n})} (u)
\label{5-8}
\eeqar
where a set of $h_i = \pm$ $(i = 1, 2, \cdots , n)$ gives an arbitrary
helicity configuration.
A similar expression is also obtained in \cite{Abe:2004ep}.

To recapitulate the results of this section, we find that the S-matrix functional
for the gluon amplitudes in general is given by
\beqar
\nonumber
\F \left[ a^{(h)c} \right] &= &  \widehat{W} \, \F_{MHV} \left[ a^{(h)c} \right] \\
&=&
\exp \left[ \int_{x, y} ~ \frac{\del_{kl}}{q^2} ~
\frac{\del}{\del a_{k}^{(+)}(x)} \otimes
\frac{\del}{\del a_{l}^{(-)}(y)} \right]\,
\exp \left[ \frac{i}{g^2} \int_{x , \th} \Theta_{R, \ga} (u; x, \th) \right]
\label{5-9}
\eeqar
where $\Theta_{R, \ga} (u; x, \th)$ is the supersymmetric holonomy operator
of gauge fields in twistor space that we have defined in the previous sections.
This expression gives a succinct realization of the CSW rules in terms of
functional derivatives.
As discussed earlier, the symbols $R$ and $\ga$ denote
the representation of the gauge group and
a loop on the physical configuration space $\C = {\bf C}^n / \S_n$, respectively.
The introduction of contraction operator in the definition of the
S-matrix functional may seem artificial but
it is a price that we have to pay for the MHV-based supersymmetric description
of gluon amplitudes.
This would also be related to our choice of ``path ordering''
in the definition of the holonomy operator in (\ref{3-28}).

\section{Concluding remarks}

In the present paper, we introduce a notion of
holonomies for gauge fields in twistor space.
This notion arises from an attempt to understand Nair's observation of
the MHV gluon amplitudes in a more universal point of view.
Nair has shown that the MHV amplitudes
can be interpreted as a current correlator of a Wess-Zumino-Witten (WZW) model
defined in a $\cp^1$ fiber of supertwistor space \cite{Nair:1988bq}.
A correlator of a WZW model generally obeys the KZ equation.
It is known that a monodromy representation of the KZ equation is given by a
linear representation of a braid group.
We take advantage of this mathematical fact to define
a holonomy operator $\Theta_{R, \ga} (u)$
in twistor space as shown in (\ref{3-28})-(\ref{3-31}).

The KZ equation has a bialgebraic structure, which naturally leads
to a bialgebraic property in what we call the comprehensive
gauge field $A$; we have constructed a holonomy operator of
this gauge field.
In a conventional field theory, where a gauge field operator
is described by creation and annihilation operators of a single particle,
we need to make a different treatment for a multi-particle systems
with different number of particles.
In the case of pure Yang-Mills theory, however,
the notion of number is not as distinctive as in theories with matter fields;
notice that gauge fields represent an interaction and, needless to say, its strength
is not determined by the number of fields
but by the value of a coupling constant.
Thus, an appropriate physical operator for gauge fields may be constructed
by summing over possible numbers of particles involved in any physical process,
such that physics with arbitrary number of particles can be described comprehensively.
This is one of the reasons for the appearance of $A$ in the description of gluons.

Another important factor in $A$ is that of the logarithmic one-form
$\om_{ij} = d \log (u_i u_j)$,
where $u_i$ and $u_j$ are spinor momenta of the $i$-th and $j$-th gluons
($1 \le i < j \le n$), respectively.
This one-form is suitable not only because the scale invariance of
$u$'s is automatic but also because
it supports a postulate of ours that in the spinor-momenta formalism
any physical observables can be described
by combinations of the Lorentz invariant scalar products
$(u_i u_j)$ and $[\bu_i \bu_j]$ defined in (\ref{2-5});
notice that if the observables are purely holomorphic, we solely need
the holomorphic products $(u_i u_j)$.
This means that the formalism is manifestly Lorentz invariant, and
hence, we can even consider this condition as a principle rather than a postulate.

As remarked earlier, the holonomy operator $\Theta_{R, \ga} (u)$
has an explicit Yangian symmetry,
while this symmetry is usually hidden in many approaches to Yang-Mills theory.
Here, the Yangian symmetry in $\Theta_{R, \ga} (u)$
is mathematically transparent if we use the Kohno-Drinfel'd monodromy theorem.
The theorem states that the monodromy representation of the KZ equation
as a linear representation of a braid group
is equivalent to a representation of the quantum Yang-Baxter equation \cite{Chari:1994pz}.
The comprehensive gauge one-form $A$, which is essentially equivalent to
the so-called KZ connection in mathematics, satisfies the flatness condition (\ref{3-19}).
Thus the holonomy operator of $A$ gives the linear representation of a braid group
$\B_n$ on the physical Hilbert space $V^{\otimes n}$, {\it i.e.},
\beq
\Theta_{R, \ga} (u) : \B_n \rightarrow GL(V^{\otimes n})
\label{6-1}
\eeq
Therefore, by use of Kohno-Drinfel'd monodromy theorem, we can easily find that
$\Theta_{R, \ga} (u)$ has the Yangian symmetry.

The braid relations in (\ref{3-2}) are closely related to the Yangian symmetry.
In fact, the only difference is the existence of the so-called spectral parameter
for the latter. As discussed earlier,
in terms of spectral parameters $\tau_i$ ($0 \le \tau_i \le 1$)
for the index $i$ ($i=1,2,\cdots, n)$,
the ``path ordering'' of $\Theta_{R, \ga} (u)$ can be
expressed as $0 \le \tau_1 \le \tau_2 \le \cdots \le \tau_n \le 1$.
In a conventional field theory, path ordering can be translated into
time ordering.
So we may interpret a set of $\tau_i$'s as a parameter for (proper) time.
On the other hand, in the construction of S-matrix functionals
for gluon amplitudes in terms of $\Theta_{R, \ga} (u)$, we find that
the path-ordering factor of $\Theta_{R, \ga} (u)$ is somehow
related to the use of supersymmetry (although we do not have
clear interpretation of this relation at the present).
In this sense, we can consider that the notion of time is
indirectly related to the requirement of supersymmetry.

We now consider some mathematical aspects of the holonomy formalism
besides what has been mentioned.
An ordinary notion of self-duality is not appropriate in the present formalism
because the holonomy operator is defined in
differential manifolds of arbitrary dimension, not just four, with
four-dimensional spacetime arising form an imposition of
the supertwistor-space condition (\ref{4-1}).
This suggests that we do not have instantons as solutions to
physical configurations of a single gluon. If we denote
a configuration space of a single gluon by ${\cal C}_{1}$, then
the existence of instantons is mathematically related to the
fact that the fundamental homotopy group of ${\cal C}_{1}$ is
given by integer, {\it i.e.},
$\Pi_1 ({\cal C}_{1}) = {\bf Z}$, for $SU(n)$ gauge groups.
(For a clear explanation of this fact, one may refer to \cite{NairBook}.)
In the present formalism, we do not consider
a single-particle configuration but a multi-particle configuration,
which we have denoted as ${\cal C} = {\bf C}^n / \S_n$, and
we should emphasize that this satisfies
\beq
\Pi_{1} ({\cal C}) = \B_n
\label{6-2}
\eeq
Thus it is reasonable to say that we do not have instantons
in our formalism but a generalization of them encoded with a braid group.

A crucial mathematical notion in the construction of holonomy operator
is the flatness or integrability of $A$; $DA = dA - A \wedge A = 0$.
As discussed before, this is the very reason why we can
define the holonomy operator itself.
A main objective of this paper is to express an S-matrix functional
for gluon amplitudes in terms of this holonomy operator.
An explicit form of the S-matrix functional is culminated in (\ref{5-9}).
This S-matrix functional naturally leads to the WZW correlator structure
for the gluon MHV amplitudes as initially observed by Nair \cite{Nair:1988bq}
and also gives an alternative perspective to Witten's generalization
of Nair's observation \cite{Witten:2003nn}.
Although the concentration has been on the gluon amplitudes in the present paper,
the holonomy operator itself is of universal nature.
We may in fact postulate that any four-dimensional integrable models
with general covariance can be generated by variants of the holonomy operator
$\Theta_{R, \ga} (u)$, possibly furnished with supersymmetry and nontrivial
Chan-Paton factors.
It is also interesting to notice that
many topics in mathematics and physics, such as conformal invariance,
Lorentz invariance, supersymmetry and Yangian symmetry,
are knit together by the holonomy operator in twistor space.

The argument on the gluon amplitudes has been limited to the classical level
but quantum descriptions seem to be straightforward since, throughout the
present paper, we consider operators that act on the quantum
Hilbert space $V^{\otimes n}$.
The S-matrix functional (\ref{5-9}) is expressed in terms of
these operators and there is nothing that prevents it
form producing loop amplitudes.
Therefore, it is possible to extend our formula to
loop diagrams without any modifications, although
confirmation of this statement is beyond the scope of the present paper.

For non-supersymmetric descriptions of gluon amplitudes, we need to define
physical operators in terms of appropriate polarization vectors
such that the degrees of homogeneity in spinor momenta
are in accord with helicities of particles.
In the holonomy formalism, conformal invariance plays an
essential role.
Thus the non-supersymmetric description may be possible only for the Abelian case.
For non-Abelian cases, supersymmetry is necessary in this respect.
Supersymmetry also has a positive or economical aspect in regard with the
issue of renormalization.

Lastly, we would like to remark that
the $N$ dependence of the coupling constant can be
written as $g = {1 \over {1 + N}}$ for $SU(N)$ gauge groups as discussed in (\ref{4-31}).
This relation shows a nonperturbative aspect of the holonomy formalism.

\vskip .3in
\noindent
{\bf Acknowledgments} \vskip .06in\noindent
The author would like to thank the Yukawa Institute for Theoretical Physics at Kyoto University.
Discussions during the YITP workshop YITP-W-08-04 on
``Development of Quantum Field Theory and String Theory'' were useful for the present work.


\end{document}